\documentclass[preprint,12pt]{elsarticle}

\usepackage{amssymb}
\usepackage{amsthm}
\usepackage{mathrsfs}
\usepackage{amsmath}
\usepackage{tabularx}
\usepackage{natbib}

\newtheorem{exmp}{Example}
\newcommand{\bs}{\boldsymbol}
\newcommand{\mb}{\mathbf}
\newcolumntype{Y}{>{\centering\arraybackslash}X} 

\journal{Compuational Statistics \& Data Analysis}

\begin{document}

\begin{frontmatter}



\title{Functional Factorial $K$-means Analysis}


 \author[MY]{Michio Yamamoto\corref{cor1}}
 \ead{michyama@kuhp.kyoto-u.ac.jp}
 \author[YT]{Yoshikazu Terada}
 \ead{terada@sigmath.es.osaka-u.ac.jp}
 \address[MY]{Department of Biomedical Statistics and Bioinformatics,
 Kyoto University Graduate School of Medicine, 54 Kawahara-cho, Shogoin,
 Sakyo-ku, Kyoto 606-8507, Japan.\\ TEL:+81-(0)75-751-4745,
 FAX:+81-(0)75-751-4732.}
 \address[YT]{Division of Mathematical Science, Graduate School of
 Engineering, Osaka University, 1-3, Machikaneyama-cho, Toyonaka, Osaka
 560-8531, Japan.\\ TEL$\setminus$FAX:+81-(0)6-6850-6490.}

 \cortext[cor1]{Corresponding author.}

\begin{abstract}
 A new procedure for simultaneously finding the optimal cluster
 structure of multivariate functional objects and finding the subspace
 to represent the cluster structure is presented. The method is based on
 the $k$-means criterion for projected functional objects on a subspace
 in which a cluster structure exists. An efficient alternating
 least-squares algorithm is described, and the proposed method is
 extended to a regularized method for smoothness of weight functions. To
 deal with the negative effect of the correlation of coefficient matrix
 of the basis function expansion in the proposed algorithm, a two-step
 approach to the proposed method is also described.  Analyses of
 artificial and real data demonstrate that the proposed method gives
 correct and interpretable results compared with existing methods, the
 functional principal component $k$-means (FPCK) method and tandem
 clustering approach. It is also shown that the proposed method can be
 considered complementary to FPCK.
 \end{abstract}


\end{frontmatter}


\section{Introduction}
\label{intro}

Cluster analysis of functional objects is often carried out in
combination with dimension reduction (e.g., Illian et al., 2009;
Suyundykov et al., 2010). In this so-called subspace clustering, a
low-dimensional representation of functional objects is used for
detecting a cluster structure of objects, rather than overall functional
objects, which may contain some irrelevant information that are likely
to hinder or completely obscure the recovery of the cluster
structure. The use of a low-dimensional representation of functional
objects can be of help in providing simpler and more interpretable
solutions.

There are two types of subspace clustering techniques: one intends to
find a subspace that is common to all clusters (Timmerman et al., 2010),
and the other intends to find a subspace specific to each cluster
(Vidal, 2011). Here, we focus on the common subspace clustering. A
frequently used approach to common subspace clustering in a functional
setting is to apply a dimension-reduction technique, such as functional
principal component analysis (FPCA) (e.g., Ramsay and Silverman, 2005;
Besse and Ramsay, 1986; Boente and Fraiman, 2000), to obtain a fewer
number of components than the overall functional data measured at a
number of time points, and subsequently to use the component scores for
clustering objects. Although it is easy to use, this two-step sequential
approach, also called the {\it tandem analysis}, provides no assurance
that the components extracted in the first step are optimal for the
subsequent clustering step, because the two steps are carried out
independently (e.g.,Arabie and Hubert, 1994; DeSarbo et al., 1990; De
Soete and Carroll, 1994; Vichi and Kiers, 2001; Timmerman et al.,
2010). In fact, each step aims to optimize a different optimization
criterion, so that tandem analysis is likely to fail in providing an
optimal cluster structure.

To overcome the problem of tandem analysis, a method that can
simultaneously perform clustering and dimension reduction is needed.
Recently, a few simultaneous procedures have been proposed. Bouveyron
and Jacques (2011) developed a model-based clustering method for
functional data that finds cluster-specific functional
subspaces. Yamamoto (2012) proposed a method, called functional
principal component $k$-means (FPCK) analysis, which attempts to find an
optimal common subspace for the clustering of multivariate functional
data. As described in Yamamoto (2012), FPCK analysis can be considered
to be an extension of the reduced $k$-means (REDKM) analysis (De Soete
and Carroll, 1994) to the model for the functional setting. Gattone and
Rocci (2012) has developed a clustering procedure that can also be
considered to be a functional version of REDKM analysis. In their
article, an efficient iteration scheme for selecting the smoothing
parameter was proposed.

Yamamoto (2012) shows that in various cases the FPCK method can find
both an optimal cluster structure and the subspace for the
clustering. The FPCK method, however, has a drawback caused by the
definition of its loss function. The drawback will be explained in more
detail in the next section. In this paper, to overcome this drawback, we
present a new method that simultaneously finds the cluster structure and
reduces the dimension of multivariate functional objects. It will be
shown that the proposed method has a mutually complementary relationship
with the FPCK model.

This paper is organized as follows. Section 2 defines the notation used
in this paper and discusses the drawbacks of FPCK analysis. In Section
3, a new clustering and dimension reduction method for functional
objects is described and an algorithm to implement the method is
proposed. In Section 4, the performance of the proposed method is
studied using artificial data, and an illustrative application to real
data is presented in Section 5. Finally, in Section 6, we conclude the
paper with a discussion and make recommendations for future research.

\section{Notation and the Drawbacks of the FPCK Method}

\subsection{Notation}

First we present the notation that we will use throughout this
paper. Here, the same notations as Yamamoto (2012) will be used for ease
of explanation. Suppose that the $n$th functional object ($n=1,\dots,N$)
with $P$ variables is represented as $x_{n}(t)=(x_{np}(t)\mid
p=1,\dots,P)$ with a domain $T\subset \mathbb{R}^{d}$. For simplicity,
we write $x_{n}=(x_{n}(t)\mid t\in T)$ to denote the $n$th observed
function. In the rest of paper, for general understanding of the
problem, we consider the single-variable case, i.e., $P=1$; in this
case, the suffix $p$ in the above notation will be omitted. The
multivariate case will be described in Appendix A. Let
$\mathscr{L}=L^{2}(T)$, which is the usual Hilbert space of function $f$
from $T$ to $\mathbb{R}$. Here, the inner product for any
$x,y\in\mathscr{L}$ is defined as
\begin{equation*}
 \left<x,y\right>:=\int_{T}x(t)y(t)dt,
	\label{eq_ffkm:inner_prod}
\end{equation*}
and for any $x\in\mathscr{L}$, $\|x\|:=\left<x,x\right>^{1/2}<\infty$.

For simplicity, we shall assume that the mean function of the $x_{n}$'s
has been subtracted, so without loss of generality, we assume that
$\sum_{n=1}^{N}x_{n}(t)=0$ for all $t\in T$.

In this paper, we simultaneously find an optimal projection of the data
$\boldsymbol{x}=(x_{1},\cdots,x_{N})'$ into a low-dimensional subspace
and a cluster structure. Let $V=\{v_{l}\}$
$(l=1,\dots,L<\infty;\;v_{l}\in\mathscr{L})$ be orthonormal basis
functions of the projected low-dimensional subspace. In this paper, as
with Yamamoto (2012), we call $v_{l}$ a weight function. In addition,
let $\mathtt{P}_{v}$ be an orthogonal projection operator from the
functional data space $\mathscr{L}$ onto the subspace $\mathscr{S}_{v}$,
which is spanned by $V$. Let $U=(u_{nk})_{N\times K}$ be cluster
assignment parameters, where $u_{nk}$ equals one if subject $n$ belongs
to cluster $k$, and zero, otherwise. Let $N_{k}$ be the number of
subjects that are assigned to the $k$th cluster, and for all $k$,
$\bar{x}_{k}:=N_{k}^{-1}\sum_{n=1}^{N}u_{nk}x_{n}$, which is the
centroid of the $k$th cluster. In this paper, we consider crisp
clustering, in which each object is assigned to only one group.

A basis function expansion approach is used in many functional
data analysis models. Let us approximate an object $x_{n}$ using
a basis function, as follows
\begin{equation*}
 x_{n} \thickapprox \sum_{m=1}^{M}g_{nm}\phi_{m}
	=\boldsymbol{\phi}'\boldsymbol{g}_{n},
\end{equation*}
where $\phi_{m}$'s $(m=1,\dots,M)$ are basis functions (e.g., Fourier
or B-spline basis functions) and $g_{nm}$ is a coefficient corresponding
to $(x_{n}, \phi_{m})$, and we write
$\boldsymbol{\phi}=(\phi_{1},\dots,\phi_{M})'$ and
$\boldsymbol{g}_{n}=(g_{n1},\cdots,g_{nM})'$. Then, we have
\begin{equation}
 (x_{1},\dots,x_{N})'\thickapprox(\boldsymbol{g}_{1},\dots,\boldsymbol{g}_{N})'\boldsymbol{\phi}
	=\mathbf{G}\boldsymbol{\phi}.
 \label{eq_ffkm:expansion_data}
\end{equation}
Similarly, the weight functions described above are expanded by the same
basis functions,
\begin{equation*}
 v_{l} \thickapprox \sum_{m=1}^{M}a_{lm}\phi_{m} =
	\boldsymbol{\phi}'\boldsymbol{a}_{l},
\end{equation*}
where $\boldsymbol{a}_{l} = (a_{l1},\cdots,a_{lM})'$. Then, we also have
 \begin{equation}
	(v_{1},\dots,v_{L})'\thickapprox (\boldsymbol{a}_{1},\dots,\boldsymbol{a}_{L})'\boldsymbol{\phi}=\mathbf{A}'\boldsymbol{\phi}.
	 \label{eq_ffkm:expansion_weightfunction}
 \end{equation}
Let $\mathbf{H}$ be an $M\times M$ matrix that has
$\left<\phi_{i},\phi_{j}\right>$ for the $ij$th
element. Furthermore, let
$\mathbf{G}_{H}=\mathbf{G}\mathbf{H}^{\frac{1}{2}}$, and
$\mathbf{A}_{H}=\mathbf{H}^{\frac{1}{2}}\mathbf{A}$.

\subsection{Drawbacks of the FPCK model}

\begin{figure}[!tb]
 \begin{center}
\renewcommand{\arraystretch}{1}
 \vspace{0.5cm}
	\begin{tabular}{cc}
	\includegraphics[width=7cm]{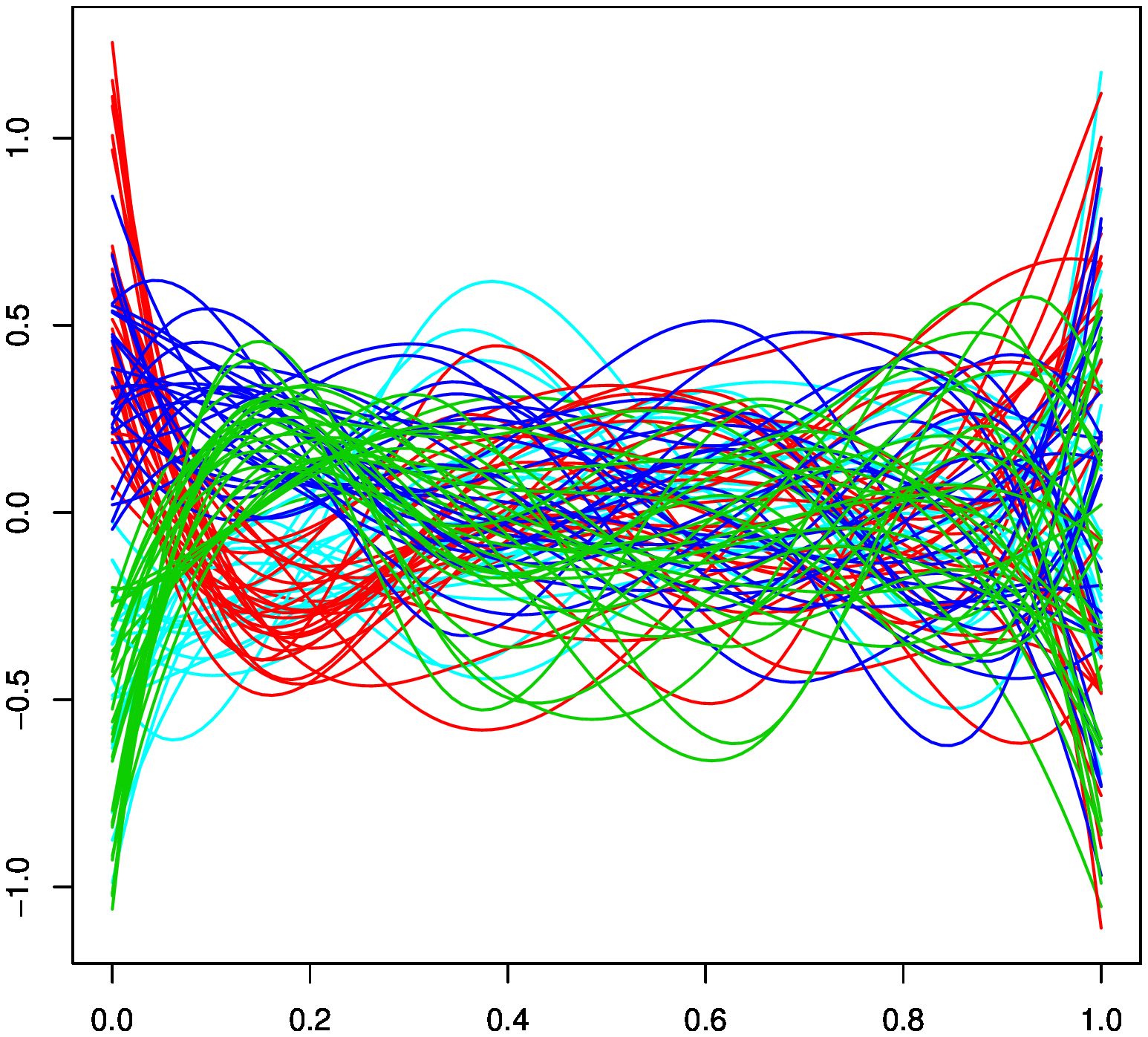} & \includegraphics[width=7cm]{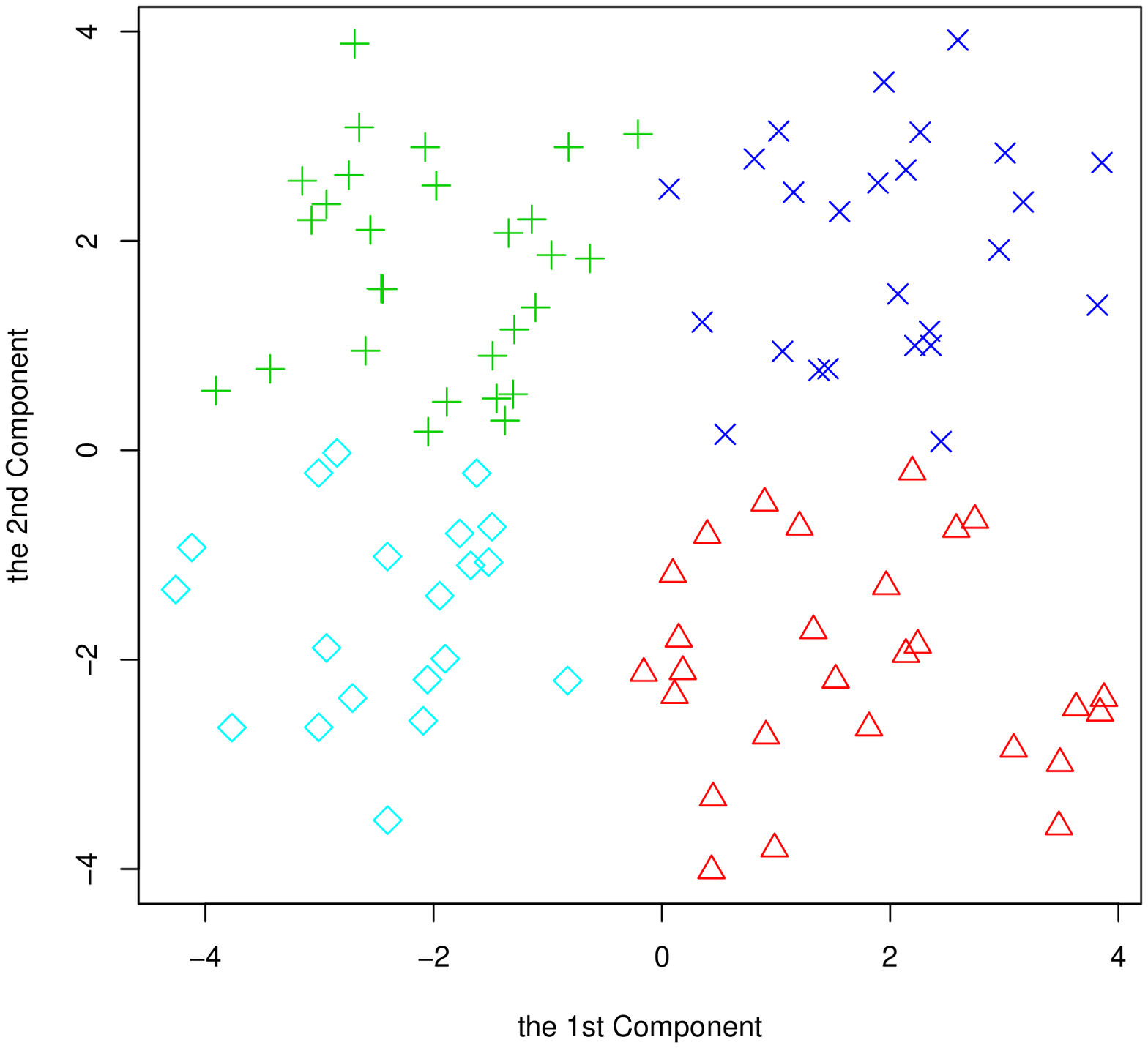}
	 \end{tabular}
 \caption{Curves of 100 functional objects (left) and the true cluster
	structure in a two-dimensional subspace (right). The colors and symbols indicate the cluster in which each object is grouped.}
	\label{fig_ffkm:intro}
\end{center}
\end{figure}

\begin{figure}[!tb]
 \begin{center}
	\renewcommand{\arraystretch}{1}
 \vspace{0.5cm}
	\begin{tabular}{c}
	\includegraphics[width=7cm]{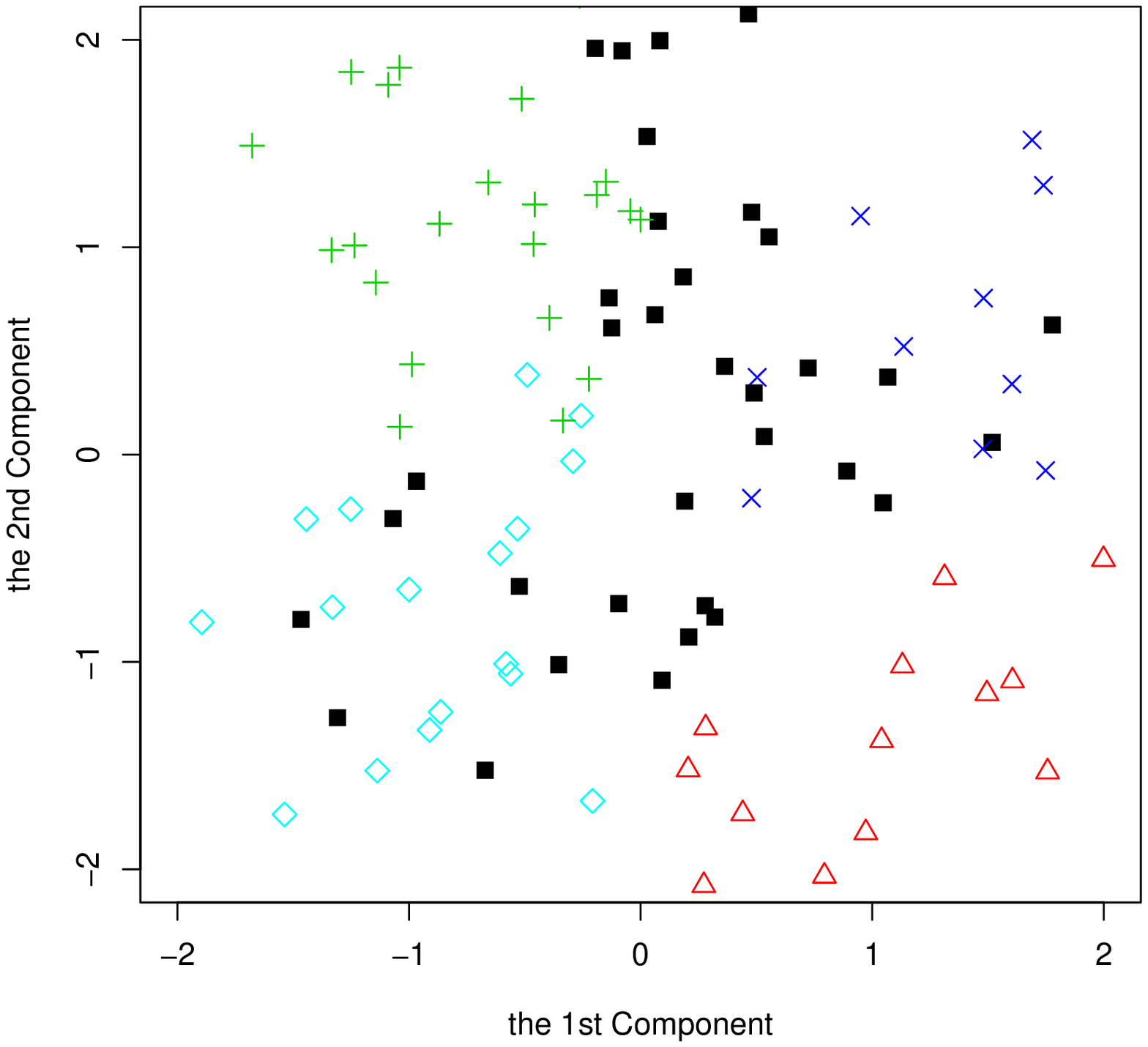}
	 \end{tabular}
 \caption{Estimated cluster structure by the FPCK method with two
	dimensions and four clusters. Colors and symbols indicate the
	cluster in which each object is grouped. A black square denotes a
	misclustered object.} \label{fig_ffkm:fpck_intro}
\end{center}
\end{figure}

As described previously, the clustering method with dimension reduction
can produce useful information about the cluster structure that exists
in functional data. To attain this purpose, the functional principal
component $k$-means (FPCK) method has been proposed (Yamamoto, 2012),
and this method succeeds in extracting a cluster structure that provides
useful information. However, the FPCK method has a drawback. A typical
example in which the FPCK analysis does not perform well is given as
follows:
\begin{exmp}
 Consider that a $100\times 6$ coefficient matrix $\mathbf{G}_{H}$
 consists of two parts,
 $\mathbf{G}_{H}=(\mathbf{G}_{1},\mathbf{G}_{2})$, where
 $\mathbf{G}_{1}$ is a $100\times 2$ matrix which defines a cluster
 structure and $\mathbf{G}_{2}$ is a $100\times 4$ matrix whose elements
 are generated randomly independent of the cluster
 structure. $\mathbf{G}_{1}$ is shown in the right of Figure
 \ref{fig_ffkm:intro}, and the left of the figure shows functional data
 of $100$ objects generated through the basis function expansion using
 $\mathbf{G}_{H}$ as its coefficient. If the FPCK method is applied to
 this data, we obtain the result shown in Figure
 \ref{fig_ffkm:fpck_intro}. As seen in Figure \ref{fig_ffkm:fpck_intro},
 the FPCK method fails to recover the true cluster structure, since
 there are many misclustered objects.
\end{exmp}

This failure of the FPCK method can be explained through the decomposition of its
loss function. The loss function $L_{fpck}$ of the FPCK method has
the following decomposition:
\begin{equation}
 L_{fpck}(U,V)=\sum_{n=1}^{N}\|x_{n}-\mathtt{P}_{v}x_{n}\|^{2}
	+\sum_{n=1}^{N}\sum_{k=1}^{K}u_{nk}\|\mathtt{P}_{v}x_{n}-\mathtt{P}_{v}\bar{x}_{k}\|^{2}.
	\label{eq_ffkm:fpck_decomposition}
\end{equation}
If we use basis function expansions of the data and weight functions,
$L_{fpck}$ is approximated as
\begin{equation*}
 L_{fpck}(U,V)\approx\|\mathbf{G}_{H}-\mathbf{G}_{H}\mathbf{A}_{H}\mathbf{A}_{H}'\|^{2}
	+\|\mathbf{G}_{H}\mathbf{A}_{H}\mathbf{A}_{H}'-\mathbf{P}_{U}\mathbf{G}_{H}\mathbf{A}_{H}\mathbf{A}_{H}'\|^{2},
\end{equation*}
where $\mathbf{P}_{U}$ is a projection matrix onto the space spanned by
the columns of $\mathbf{U}=(u_{nk})$. The first term of the right-hand
side measures the distance between the coefficient matrix
$\mathbf{G}_{H}$ and the projection of $\mathbf{G}_{H}$ onto the
subspace spanned by the columns of $\mathbf{A}_{H}$. That is, this term
determines the degree of the dimension reduction of the data. On the other
hand, the second term measures the distance between the projection of
$\mathbf{G}_{H}$ and the centroid of clusters in the subspace. Based on
this formulation, it is found that there are some cases where FPCK analysis does
not work well. We illustrate this using a concrete example.

As with Example 1, consider that an $N\times M$ coefficient matrix
$\mathbf{G}_{H}$ consists of two parts,
$\mathbf{G}_{H}=(\mathbf{G}_{1},\mathbf{G}_{2})$, where $\mathbf{G}_{1}$
is an $N\times M_{1}$ matrix that is related to the cluster structure,
and $\mathbf{G}_{2}$ is an $N\times M_{2}$ matrix ($M=M_{1}+M_{2}$) that
is independent of the cluster structure. Usually, $N$ denotes the sample
size, and $M$ is the number of basis functions. If $\mathbf{G}_{1}$ has
no substantial correlations, then FPCK analysis is likely to provide a
different subspace from that spanned by the true $\mathbf{A}_{H}$. This
is mainly because $\mathbf{G}_{1}$ is full rank, and the first term
of the decomposition may be minimized by weight functions which
are different from true ones. It can be inferred that when
$\mb{G}_{1}$ is full rank, the FPCK method gets worse with an increase
in the column size of $\mathbf{G}_{2}$. Evidently, it can be seen that,
if the contributing part $\mathbf{G}_{1}$ to the cluster structure has
no substantial correlations and the masking part $\mathbf{G}_{2}$
substantially exists, the FPCK method may fail to find the true cluster
structure.

\section{Proposed Method}
\subsection{Criterion of the functional factorial $k$-means method}

To overcome the drawback of FPCK analysis discussed above, we propose a
new clustering method with dimension reduction. The notation and
settings were explained in Section 2. For ease of explanation, we first
consider the case in which there is only one variable, i.e.,
$P=1$. Thus, in this section, the suffix $p$ is omitted from the
notation. An extension to the multivariate model is straightforward and
is described in Appendix A.

A least-squares objective function for the proposed approach, in which
the first few principal components of the data are defined to be the most
informative about the cluster structure, is
\begin{equation}
L_{ffkm}(U,V) = \sum_{n=1}^{N}\sum_{k=1}^{K}u_{nk}
 \|\mathtt{P}_{v}x_{n}-\mathtt{P}_{v}\bar{x}_{k}\|^{2}.
 \label{eq_ffkm:target}
\end{equation}
This criterion is optimized over the projected space $V$ and the cluster
parameter $U$.

Here, a component score $f_{nl}$ of subject $n$ for the $l$th component
is defined as $f_{nl} = \langle x_{n},v_{l}\rangle$ using the estimated
weight function $v_{l}$. Analysis for the first few estimated component
scores $\{f_{nl}\}$ $(l=1,\dots,L)$, where $L$ is two or three, seems to
be helpful for the interpretation of a cluster structure in functional
data.

\begin{figure}[!htb]
 \begin{center}
	\renewcommand{\arraystretch}{1}
 \vspace{0.5cm}
	\begin{tabular}{c}
	\includegraphics[width=7cm]{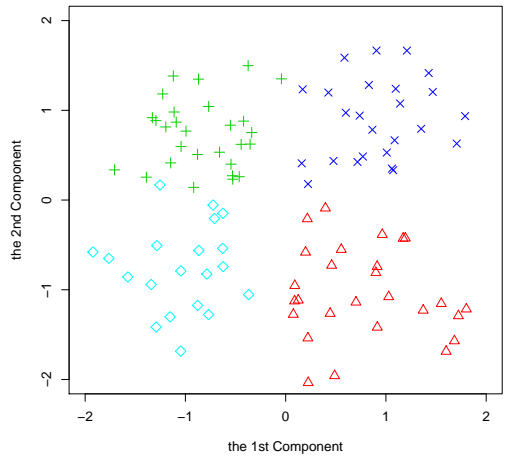}
	 \end{tabular}
 \caption{Estimated cluster structures by the FFKM method with two
	dimensions and four clusters. There were no misclustered objects.}
	\label{fig_ffkm:ffkm_intro}
\end{center}
\end{figure}

This approach, minimizing the objective function in
(\ref{eq_ffkm:target}) with respect to $V$ and $U$ simultaneously, is
called the functional factorial $k$-means (FFKM) method because this
method is a direct extension of the factorial $k$-means method (Vichi
and Kiers, 2001) to the model for the functional setting. The loss
function (\ref{eq_ffkm:target}) is equivalent to the second term of the
decomposition (\ref{eq_ffkm:fpck_decomposition}) of the loss function of
FPCK. It might be expected that we can resolve the problem of FPCK by
ruling out the first term in
Eq. (\ref{eq_ffkm:fpck_decomposition}). Note that this loss function
(\ref{eq_ffkm:target}) was shortly referred in Yamamoto (2012).
\begin{exmp}
 The FFKM method was applied to the data in Example 1. Figure
 \ref{fig_ffkm:ffkm_intro} shows the two-dimensional representation of
 the data given by FFKM. It is found that FFKM recovered the true
 cluster structure completely.
\end{exmp}

\subsection{Algorithm for optimizing the proposed criterion}
\label{sec_ffkm:algorithm} We now present an efficient algorithm for this
approach. As in the FPCK method, the criterion
(\ref{eq_ffkm:target}) can be optimized using the alternating least squares
(ALS) approach, as follows.
\begin{enumerate}
 \setlength{\itemindent}{30pt}
 \item[{\it STEP1}.] Initialize parameter $V$ subject to the restriction
								 mentioned above.
 \item[{\it STEP2}.] Minimize the loss function in
								 Eq. (\ref{eq_ffkm:target}) for fixed $V$ over $U$.
 \item[{\it STEP3}.] Minimize the loss function in Eq. (\ref{eq_ffkm:target})
								 for fixed $U$ over $V$.
 \item[{\it STEP4}.] Go to {\it STEP2}, or stop.
\end{enumerate}
There are two parts to the algorithm. The first part of the above ALS
algorithm is to minimize $L_{ffkm}$ for fixed $V$ over $U$. To solve the
optimization problem, we use a basis function expansion technique
described in Section 2. If a projected object $\mathtt{P}_{v}x_{n}$ is
expanded using some basis function, that is,
$\mathtt{P}_{v}x_{n}=\boldsymbol{d}'_{n}\boldsymbol{\phi}$ where
$\boldsymbol{d}_{n}=(d_{n1},\dots,d_{nM})'$, then the criterion
(\ref{eq_ffkm:target}) can be written as
\begin{equation}
\sum_{n=1}^{N}\sum_{k=1}^{K}u_{nk}\|\mathtt{P}_{v}x_{n}-\mathtt{P}_{v}\bar{x}_{k}\|^{2}
	= \sum_{n=1}^{N}\sum_{k=1}^{K}u_{nk}\|\boldsymbol{d}_{n}-\bar{\boldsymbol{d}}_{k}\|_{\mathbf{H}}^{2},
	\label{eq_ffkm:k-means_algorithm}
\end{equation}
where $\bar{\boldsymbol{d}}_{k}$ is a coefficient vector corresponding
to the basis function expansion of the projected mean function
$\bar{x}_{k}$ of the $k$th cluster, and $\|\cdot\|_{\mathbf{H}}$ means
the Euclidean norm with the metric $\mathbf{H}$, i.e., for
$\mathbf{y}\in \mathbb{R}^{M},\;
\|\boldsymbol{y}\|_{\mathbf{H}}^{2}=\boldsymbol{y}'\mathbf{H}\boldsymbol{y}$. Thus,
Eq. (\ref{eq_ffkm:k-means_algorithm}) can be minimized using the usual
$k$-means algorithm (Lloyd, 1982) for
$\mathbf{H}^{\frac{1}{2}}\boldsymbol{d}_{n}$. Using the expansions in
Eq. (\ref{eq_ffkm:expansion_weightfunction}), it is found that
$\mathbf{H}^{\frac{1}{2}}\boldsymbol{d}_{n}=\mathbf{A}_{H}\mathbf{A}_{H}'\boldsymbol{g}_{Hn}$.

The second part is to minimize $L_{ffkm}$ regarding $V$. The objective
function in Eq. (\ref{eq_ffkm:target}) can be written as (see Yamamoto,
2012, p.246)
\begin{align}
 L_{ffkm}(U,V) &=-\sum_{l=1}^{L}\langle v_{l},\mathtt{F}v_{l}\rangle,
 \label{eq_ffkm:algorithm}
\end{align}
where $\mathtt{F}$ is an integral operator defined as, for any
$y\in\mathscr{L}$,
\begin{align*}
 \label{eq_ffkm:func_model_eq_F}
 (\mathtt{F}y)(t) &:= -\sum_{n=1}^{N}\sum_{k=1}^{K}u_{nk}
 \langle x_{n}-\bar{x}_{k},y\rangle
 (x_{n}(t)-\bar{x}_{k}(t)).
\end{align*}
Note that it is easily verified that the integral operator $\mathtt{F}$
is a Hilbert-Schmidt integral operator. Thus, $\mathtt{F}$ is a compact
operator. In addition, $\mathtt{F}$ is clearly self-adjoint. Minimizing
the criterion is, therefore, equivalent to solving the following
eigenvalue equation (see, for example, Dunfort and Schwartz, 1988),
\begin{equation}
 \mathtt{F}\mathbf{\xi}_{l}=\rho_{l}\mathbf{\xi}_{l},\hspace{10pt}\text{
	subject to}\hspace{10pt}\left<\xi_{l},\xi_{l'}\right>=\delta_{ll'}
	\label{eq_ffkm:eigen_prob}
\end{equation}
for $l=1,\dots,L$, where $\delta_{ll'}$ is the Kronecker delta. Each
eigenfunction $\{\xi_{l}\}$ $(l=1,\dots,L)$ corresponds to a weight
function $\{v_{l}\}$ $(l=1,\dots,L)$, which is to be estimated. As with
the first part of the ALS algorithm, to solve this eigenvalue problem,
we use the basis function expansion. Then, $\mathtt{F}$ operates on a
function $\xi_{l}$ as
\begin{align*}
 (\mathtt{F}\xi_{l})(t) =\boldsymbol{\phi}'(t)\mathbf{G}'(\mathbf{P}_{U}-\mathbf{I}_{N})\mathbf{G}\mathbf{H}\boldsymbol{a}_{l}.
\end{align*}
Eventually, solving the eigenvalue problem (\ref{eq_ffkm:eigen_prob})
amounts to solving the eigenvalue problem
\begin{equation*}
 \mathbf{G}_{H}'(\mathbf{P}_{U}-\mathbf{I}_{N})\mathbf{G}_{H}\boldsymbol{a}_{Hl}=\rho\boldsymbol{a}_{Hl},
\end{equation*}
where
$\boldsymbol{a}_{Hl}=\mathbf{H}^{\frac{1}{2}}\boldsymbol{a}_{l}$. The
eigenfunction $\xi_{l}$ is given by the estimated eigenvector
$\boldsymbol{a}_{Hl}$ as the approximation in
Eq. (\ref{eq_ffkm:expansion_weightfunction}) using
$\boldsymbol{a}_{l}=\mathbf{H}^{-\frac{1}{2}}\boldsymbol{a}_{Hl}$.

The above ALS algorithm monotonically decreases the loss function
$L_{ffkm}$ and the loss function is bounded from below. Then this
algorithm guarantees the convergence to a certain point; but it may not
be the global minimum. Also, in general, the $k$-means algorithm, which
is utilized in the ALS algorithm, is sensitive to local optima
(Steinley, 2003). Thus, to safeguard against those local minima, the
proposed algorithm needs to be repeated with a number of random initial
starts for $V$.

\subsection{Regularized model}
\label{sec_ffkm:penalized}

In this section, we propose a smoothing method for the FFKM
model. Generally, if the functional data can be assumed to be
sufficiently smooth, the analysis method considering the smoothness of
functions often provides better results (see, for example, Ramsay and
Silverman, 2005). Specifically, several smoothing approaches to FPCA
have been developed (Rice and Silverman, 1991; Silverman, 1996; Reiss
and Ogden, 2007) and investigated theoretically (Pezzulli and Silverman,
1993; Silverman, 1996; Ocana et al., 1999; Reiss and Ogden, 2007), from
the beginning during the early stages of research on functional data
analysis.

In this section, the FFKM method is extended to the regularized model
which takes into consideration the smoothness of functional objects and
weight functions. Here a regularized model for univariate is described,
while the method for a multivariate case is described in Appendix A. The
proposed approach to the regularized FFKM model is similar to the
approach in penalized FPCA proposed by Silverman (1996).

Let $\mathtt{D}^{2}$ be the second-order differential operator, and let
$\mathtt{S}^{2}_{\lambda}$ be the usual spline smoothing operator (see,
Green and Silverman, 1994) with a roughness penalty $\lambda$. That is,
for any function $f\in \mathscr{L}$, the loss function
$\|f-g\|^{2}+\lambda\|\mathtt{D}^{2}g\|^{2}$ with the penalty of
function $g$ is minimized when setting $g=\mathtt{S}_{\lambda}^{2}f$. We
consider the inner product space
$(\mathscr{L},\langle\cdot,\cdot\rangle_{\lambda})$ with the inner
product $\left<\cdot,\cdot\right>_{\lambda}$ which is defined as, for
$x,y\in\mathscr{L}$,
\begin{equation*}
 \langle x,y\rangle_{\lambda}:=\langle x,y\rangle + \lambda\langle\mathtt{D}^{2}x,\mathtt{D}^{2}y\rangle.
\end{equation*}
Note that the norm $\|\cdot\|_{\lambda}$ is given by the inner product,
i.e., $\|x\|_{\lambda}=\langle x,x\rangle_{\lambda}^{1/2}$. Then,
smoothed weight functions $V$ can be obtained by the FFKM method on the
smoothed functional data $\mathtt{S}_{\lambda}^{2}x_{n}\;
(n=1,\dots,N)$. Thus, a loss function of the regularized FFKM method is
\begin{equation*}
 L_{ffkm}(U,V)=\sum_{n=1}^{N}\sum_{k=1}^{K}u_{nk}\|\mathtt{P}_{v}\mathtt{S}_{\lambda}^{2}x_{n}-\mathtt{P}_{v}\mathtt{S}_{\lambda}^{2}\bar{x}_{k}\|^{2}_{\lambda}.
\end{equation*}

The parameters $U$ and $V$, which minimize $L_{ffkm}(U,V)$, are
estimated using an ALS algorithm similar to that for the non-regularized
FFKM method, though there are two differences between the two models: in
the regularized model, the inner product
$\left<\cdot,\cdot\right>_{\lambda}$ is used and the smoothed data
$\mathtt{S}_{\lambda}x_{n}$ is expanded.  Let
$\boldsymbol{g}_{\lambda,n}$ be a vector with length $M$ containing
coefficients corresponding to the basis function expansion of
$\mathtt{S}_{\lambda}^{2}x_{n}$, and let $\mathbf{H}_{\lambda}$ be an
$M\times M$ matrix in which the $ij$th element is
$\langle\phi_{i},\phi_{j}\rangle_{\lambda}$. Furthermore, let
$\mathbf{A}_{H_{\lambda}}=\mathbf{H}_{\lambda}^{\frac{1}{2}}\mathbf{A}$
and
$\mathbf{W}=\mathbf{H}_{\lambda}^{\frac{1}{2}}\mathbf{A}_{H_{\lambda}}\mathbf{A}_{H_{\lambda}}'\mathbf{H}_{\lambda}^{\frac{1}{2}}$. Then,
in {\it STEP2} of the ALS algorithm, the optimal $U$ is obtained by
minimizing the following criterion for fixed $V$ over $U$:
\begin{equation*}
 \sum_{n=1}^{N}\sum_{k=1}^{K}u_{nk}\|\boldsymbol{g}_{\lambda,n}-\bar{\boldsymbol{g}}_{k}\|_{\mathbf{W}}^{2},
\end{equation*}
where $\|\cdot\|_{\mathbf{W}}$ is the Euclidean norm with metric
$\mathbf{W}$. Thus, as with the non-regularized model, this criterion
will be optimized using the usual $k$-means algorithm for
$\mathbf{A}_{H_{\lambda}}\mathbf{A}_{H_{\lambda}}'\boldsymbol{g}_{H_{\lambda}n}$.

Next, we describe how to estimate the weight functions $V$ in {\it
STEP3}. Using the above basis function expansion to estimate the optimal
$V$, the following eigenvalue problem is considered:
\begin{equation*}
 \mathbf{G}_{H_{\lambda}}'(\mathbf{P}_{U}-\mathbf{I}_{n})\mathbf{G}_{H_{\lambda}}\boldsymbol{a}_{H_{\lambda}l}=\rho\boldsymbol{a}_{H_{\lambda}l},
\end{equation*}
where $\boldsymbol{a}_{H_{\lambda}l}$ is the $l$th column of
$\mathbf{A}_{H_{\lambda}}$. Then, as in the non-regularized method, the
smoothed weight function $v_{l}$ is approximated as
$v_{l}\approx\boldsymbol{\phi}'\mathbf{H}_{\lambda}^{-\frac{1}{2}}\boldsymbol{a}_{H_{\lambda}l}$.

A component score $f_{nl}$ can be defined as that for the FFKM method,
$f_{nl}=\langle x_{n},v_{l}\rangle$. Before the regularized FFKM method
is applied to the data, the value of the smoothing parameter $\lambda$
should be determined. Some general remarks about the use of automatic
methods for choosing smoothing parameters are found in Green and
Silverman (1994) and Wahba (1990). A more detailed explanation of
determining the value of $\lambda$ is presented in the next section.

\subsection{Model selection}
\label{sec_ffkm:model_selection}

Prior to applying the above algorithm, we need to determine the values
of parameters: the smoothness of the basis functions, the number of
clusters, and the dimensionality of the subspace. Here, we discuss in
detail how these selections should be made.

First, we discuss the selection of the smoothness of the function. As
described in Ramsay and Silverman (2005), it is often adequate for many
purposes to choose the smoothing parameter subjectively. On the other
hand, selecting the value of $\lambda$ in an automatic manner may be
required if there is no prior information on the smoothness. There are
two major approaches to the automatic way: one is to minimize the
predictive errors of the parameters specific to each problem (Silverman,
1996) and the other is to minimize the predictive errors of the curve
estimation (Kneip, 1994). In this paper, for simplicity, we adopt the
latter approach. In order to reduce computational costs, we used the
generalized cross-validation procedure to decide the value of
$\lambda$. That is, for the multivariate discrete sample $x_{npt}$
$(n=1,\dots,N; p=1,\dots,P; t=1,\dots,T)$, we used the values that
minimized the following criterion:
\begin{equation*}
 GCV(\lambda) = \sum_{n=1}^{N}\sum_{p=1}^{P}\frac{\sum_{t=1}^{T}(x_{npt}-\hat{x}_{npt})^{2}}{T\left(1-\frac{\mathrm{trace}(\mathbf{\Gamma}_{\lambda})}{T}\right)^{2}},
\end{equation*}
where $\mathbf{\Gamma}_{\lambda}$ denotes a hat matrix such that for
$\boldsymbol{x}_{np}=(x_{np1},\cdots,x_{npT})'$,
\begin{equation*}
 \hat{\boldsymbol{x}}_{np}=\mathbf{\Gamma}_{\lambda}\boldsymbol{x}_{np}.
\end{equation*}

Next, we discuss the selection of the number of clusters and the
dimensionality. For ease of explanation, we consider the univariate
case. The following discussion is also valid for the multivariate
case. The basis function expansions of the functional objects
$x_{n}$ and the weight functions $v_{l}$ provide an approximation of the
loss function as follows:
\begin{equation}
 L_{ffkm}(\mathbf{U},\mathbf{A}_{H}) \approx
	\|\mathbf{G}_{H}\mathbf{A}_{H}-\mathbf{P}_{U}\mathbf{G}_{H}\mathbf{A}_{H}\|^{2},
	\label{eq_ffkm:approximation_modelselection}
\end{equation}
where the norm is the Frobenius norm. As mentioned previously, since the
data $x_{n}(t)$ are centered at each time point $t$, the coefficient
matrix $\mathbf{G}_{H}$ is also a column-wise centered matrix. Then, the
rank of $\mathbf{P}_{U}\mathbf{G}_{H}\mathbf{A}_{H}$ is equal to or less
than $\min(K-1, L)$. Thus, the choices for the number of clusters and
components should not be made independent of each other. This situation
is the same as that in the factorial $k$-means method (Vichi and Kiers,
2001). According to the recommendation made by Vichi and Kiers (2001),
we first choose the number of clusters, and then verify an adequacy of
the dimensionality used for the analysis by checking whether the
coordinates of the cluster centroids
$(\mathbf{U}'\mathbf{U})^{-1}\mathbf{U}'\mathbf{G}_{H}\mathbf{A}_{H}$,
can be adequately represented by fewer components. To select the number
of clusters, it can be done either on the basis of subjective
information or by applying some decision procedure, such as that
described in Milligan and Cooper (1985) or Hardy (1996). For the
selection of the dimensionality, it is recommended to first take $L=K-1$
and then check the adequacy of the dimensionality. For instance, it may
be useful to check whether the cluster centroids appear to lie in a
lower-dimensional plane, in which case it is advised to refit the FFKM
model with fewer components. By thus verifying the solutions for
different numbers of clusters, one can select the solution that gives
the most interpretable results.

\section{Analyses of Artificial Data}

\subsection{Data and evaluation procedures}

To investigate the performance of the FFKM method, artificial data,
which included a known low-dimensional cluster structure, were analyzed
by four different methods: (i) the FFKM method, (ii) the two-step FFKM
method (FFKMts) (iii) the FPCK method, and (iv) tandem analysis (TA)
that consisted of FPCA using a basis function expansion (Ramsay and
Silverman, 2005) followed by a standard $k$-means cluster analysis of
the object scores on the first $L$ principal components. Note that the
loss function of FFKM is bounded above by the squared norm of the
projected functional data as follows:
\begin{equation}
 	\sum_{n=1}^{N}\sum_{k=1}^{K}u_{nk}\|\mathtt{P}_{v}x_{n}-\mathtt{P}_{v}\bar{x}_{k}\|^{2}
	 \le\sum_{n=1}^{N}\|\mathtt{P}_{v}x_{n}\|^{2}.
	\label{eq_ffkm:inequality}
\end{equation}
Thus, when an empirical covariance operator of functional data has
excessively small eigenvalues compared with the others, the subspace
spanned by eigenfunctions corresponding to the small eigenvalues
provides the smallest values of loss function of FFKM regardless of
cluster assignments. In fact, when the smallest eigenvalue of an
empirical covariance operator is zero, using the corresponding
eigenfunction as a weight function for $\mathtt{P}_{v}$ sets the value
of right-hand side of (\ref{eq_ffkm:inequality}) to zero, and then the
loss of FFKM is also zero. That is, if there exist trivial dimensions of
functional data, FFKM may fail to find the optimal cluster structure.
Thus, to avoid such trivial solutions of FFKM, here we introduce a
two-step approach, called two-step FFKM. The two-step FFKM method is a
two-step approach in which first we eliminate trivial dimensions from
the data and then apply the FFKM algorithm to the reduced data. This
two-step approach can improve the efficiency of the FFKM method when the
coefficient matrix $\mb{G}_{H}$ has some correlations. This two-step
approach is described in Appendix B in more detail. The artificial
functional data had a structure of four clusters in a two-dimensional
subspace, i.e., $L=2$ and $K=4$.

As described in Section 2.2, we suppose that the coefficient matrix
$\mathbf{G}_{H}$ consists of two parts,
$\mathbf{G}_{H}=(\mathbf{G}_{1},\mathbf{G}_{2})$, where $\mathbf{G}_{1}$
is an $N\times M_{1}$ matrix that is related to the cluster structure
and $\mathbf{G}_{2}$ is an $N\times M_{2}$ matrix that is independent of
the cluster structure. Let an $N\times L$ component score matrix
$\mathbf{F}$ have a cluster structure with $N$ objects drawn from four
bivariate normal distributions with the same covariance matrices,
$\mathbf{I}_{2}$, and different means. Let $\mathbf{A}_{1}$ be an
$M_{1}\times L$ orthonormal matrix whose elements were randomly
generated and subsequently orthonormalized. Using these matrices, the
matrix $\mathbf{G}_{1}$ was calculated as
$\mathbf{G}_{1}=\mathbf{F}\mathbf{A}_{1}'$. The elements of $\mb{G}_{2}$
were generated according to a strategy described later.

\begin{figure}[!tb]
 \begin{center}
	\begin{tabular}{c}
	\includegraphics[width=7cm]{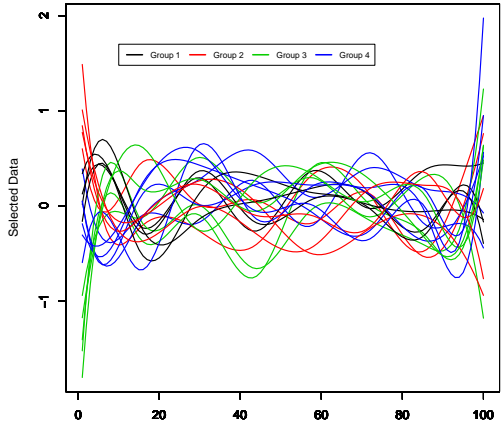}
	\end{tabular}
 \caption{Selected artificial data; the color denotes to which group each
	functional object was assigned; the proportion of overlap is 0.05.}
	\label{fig_ffkm:selected_artificial_data}
 \end{center}
\end{figure}

Let $\boldsymbol{\phi}$ be the fourth-order B-spline basis functions
with eight knots, and let $\mathbf{\Phi}$ be a $T\times M$ matrix whose
$tm$th element is $\phi_{m}(t)$. In this simulation study, we consider
100 sampling points $\boldsymbol{t}=(1,\dots,100)$ and 10 basis
functions. Then, an artificial data matrix that includes discretized
functional data was calculated as
$\mathbf{X}=\mathbf{G}_{H}\mathbf{H}^{-\frac{1}{2}}\mathbf{\Phi}'$. Note
that before calculating $\mathbf{X}$, the columns of $\mathbf{G}_{H}$
were standardized. The artificial data selected are shown in Figure
\ref{fig_ffkm:selected_artificial_data}.

\begin{figure}[!tb]
 \begin{center}
	\begin{tabular}{cc}
	 PO = 0.0001 & PO = 0.05 \\
	 \includegraphics[width=5cm,clip]{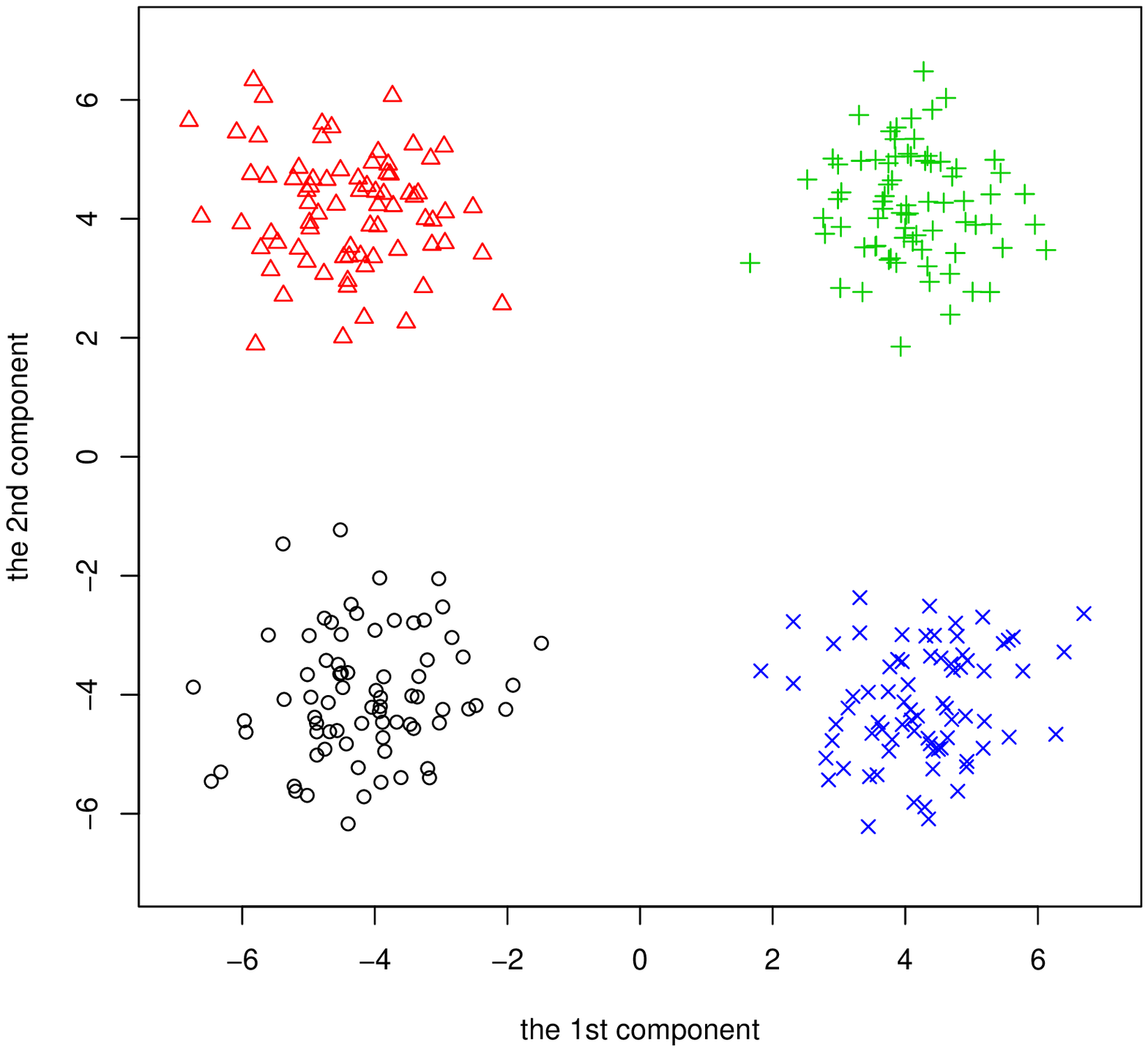} &
	 \includegraphics[width=5cm,clip]{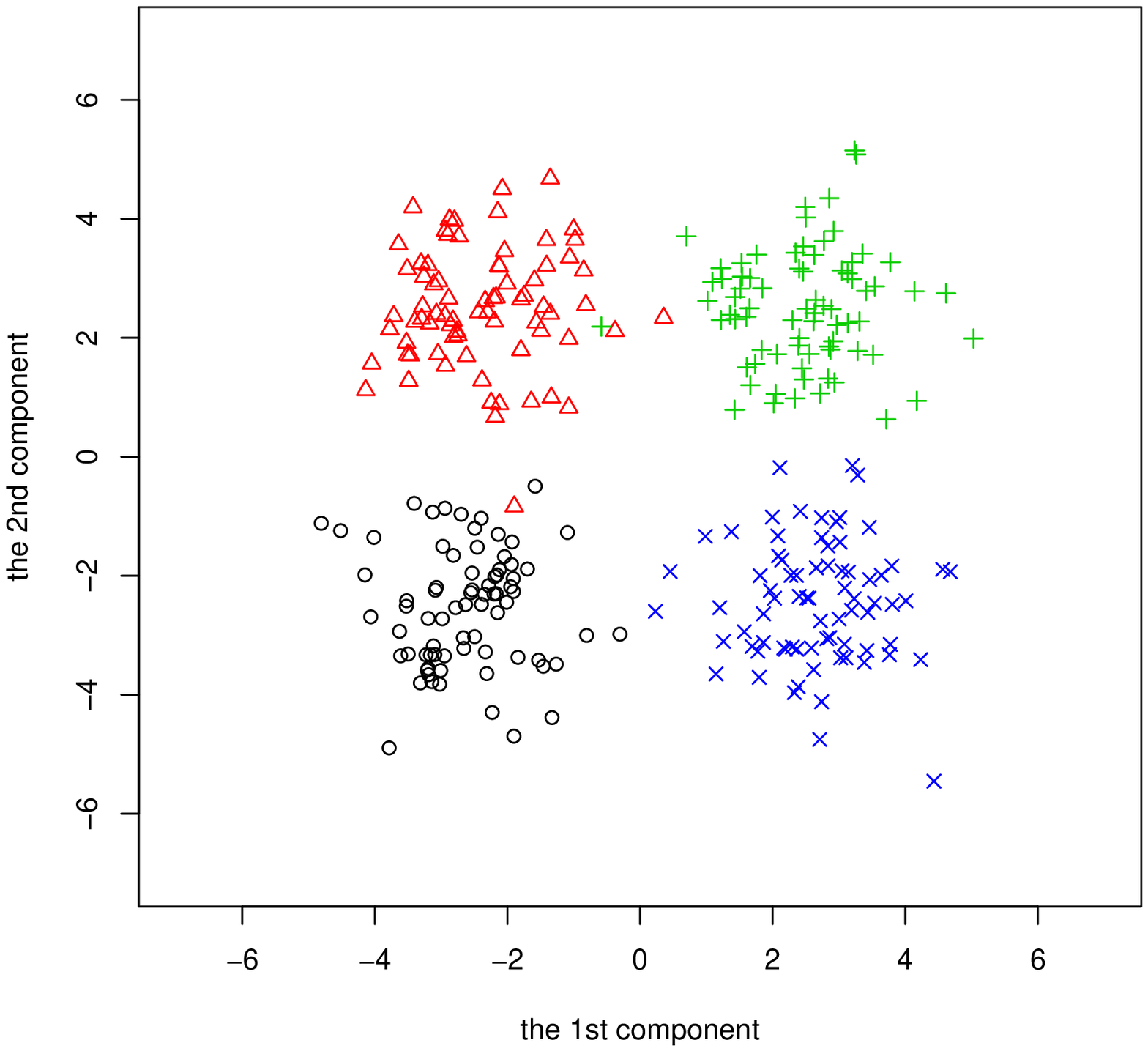}\\ PO =
	 0.10 & PO = 0.15 \\
	 \includegraphics[width=5cm,clip]{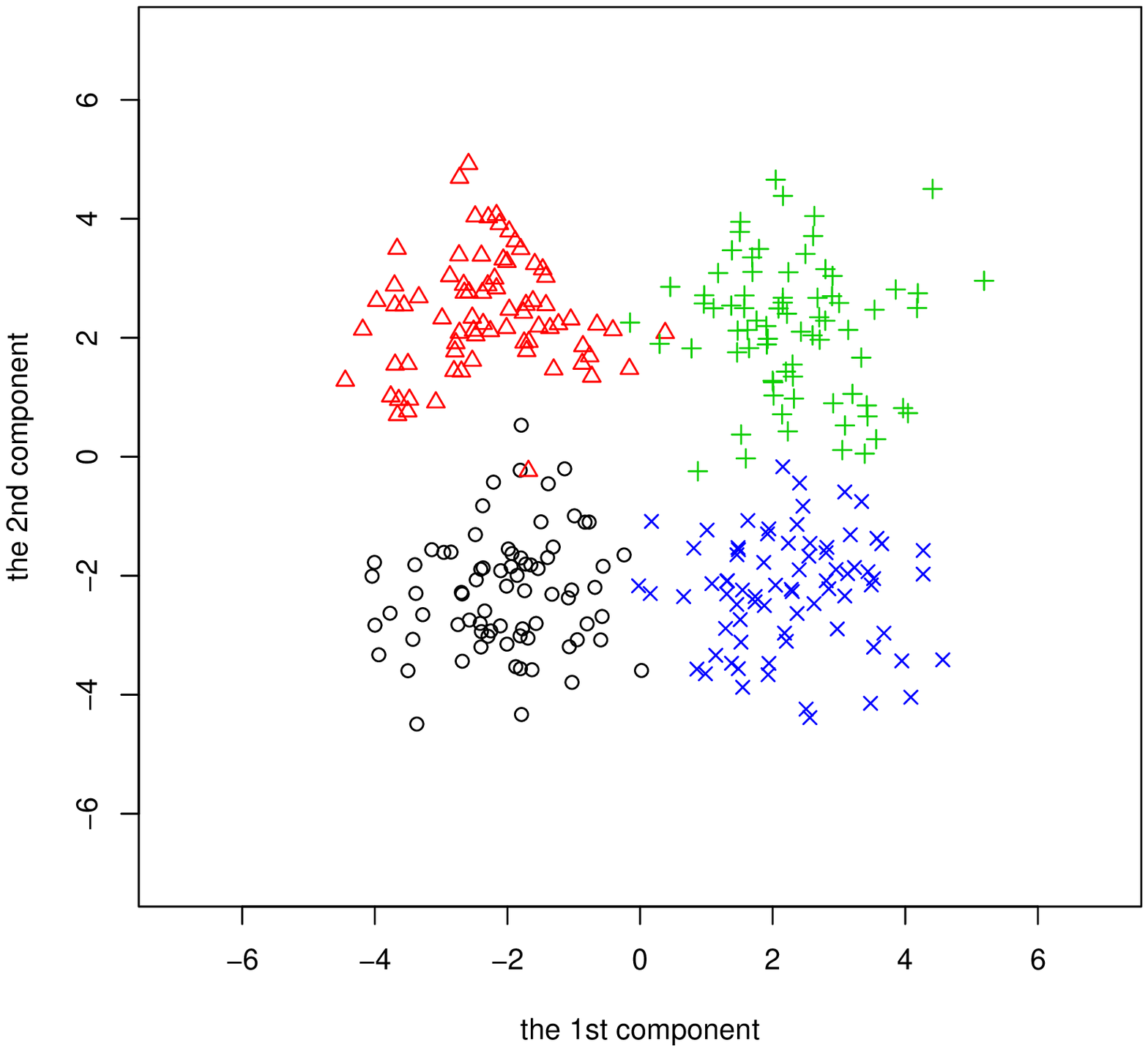} &
	 \includegraphics[width=5cm,clip]{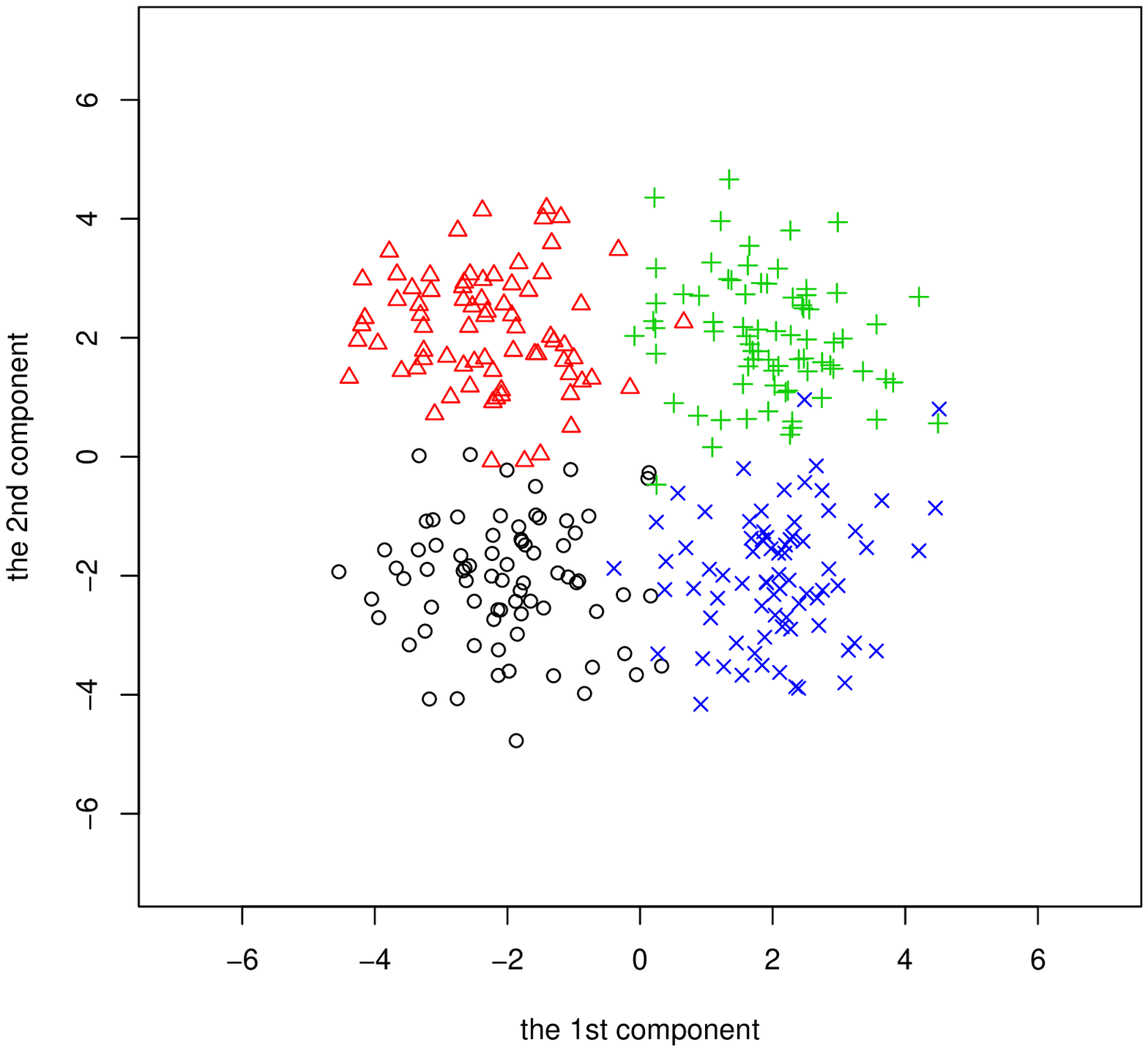}
	\end{tabular}
 \caption{Example of simulated component scores for 200 objects in four
	clusters in the correct two-dimensional subspace at four levels of
	proportion of overlap (PO)} \label{fig_ffkm:example_object_scores}
	\end{center}
\end{figure}

In this simulation analysis, four factors were manipulated in the
experiment: (1) the number of objects ($N$), (2) the expected proportion
of overlap (PO) between clusters in the correct subspace, (3) the ranks
of the coefficient matrices $\mb{G}_{1}$ and $\mb{G}_{2}$, and (4) the
number of variables which have no information about the true cluster
structure (the number of non-informative variables, NN). The number of
objects was varied from $100$ to $500$ in steps of $200$. The PO was
defined as the proportion of shared density between clusters, as
proposed by Steinley and Henson (2005). The PO was set at four levels:
0.0001, 0.05, 0.10, and 0.15. To offer an impression of the effect of
the manipulation of the PO, an example of $\mathbf{F}$ for 200 objects
in four clusters is depicted in Figure
\ref{fig_ffkm:example_object_scores}, for each of the different levels
of the PO.
We consider four cases with the combination of ranks of $\mb{G}_{1}$
and $\mb{G}_{2}$; for each coefficient matrix, we consider two cases,
full rank (FR) or rank deficient (RD). The rank of $\mb{G}_{1}$ was
controlled by the number of columns $M_{1}$, which was set at 2 for the
FR case and 5 for the RD case. For a FR case of $\mb{G}_{2}$, the
elements of $\mb{G}_{2}$ were independently drawn from a standard normal
distribution $N(0, 1)$, while for a RD case, $\mb{G}_{2}$ was calculated
as $\mb{G}_{2}=\mb{E}\mb{A}_{2}'$, where $\mb{E}$ is an $N\times
(M_{2}-2)$ matrix and $\mb{A}$ is an $M_{2}\times (M_{2}-2)$ matrix. The
size $M_{2}-2$ implies that the rank of $\mb{G}_{2}$ is two lower than
that of the FR case. The elements of $\mb{E}$ and $\mb{A}_{2}$ were
independently drawn from $N(0,1)$ and $\mb{A}_{2}$ was subsequently
orthonormalized. When $\mb{G}_{1}$ and $\mb{G}_{2}$ are FR, FFKM works
well but FPCK does not. On the other hand, when $\mb{G}_{1}$ is RD and
$\mb{G}_{2}$ is FR, FPCK works well but FFKM does not. Furthermore, it
can be inferred that both FFKM and FPCK are effected negatively by the
rank deficiency of $\mb{G}_{2}$. A non-informative variable $\mb{Z}$ was
also generated through the basis function expansion
$\mb{Z}=\mb{G}_{H}^{*}\mb{H}^{-\frac{1}{2}}\mb{\Phi}'$ in which elements
of a coefficient matrix $\mb{G}_{H}^{*}$ were independently drawn from
$N(0,1)$ and standardized to have a same variance with the informative
data $\mb{X}$. In this study, the number of non-informative variables
was set at three levels: 0, 1, and 2. The experimental design was fully
crossed, with $50$ replicates per cell, yielding
$3\times4\times4\times3\times50=7200$ simulated data sets.

The cluster membership recovery was assessed by the adjusted Rand index
(ARI; Hubert and Arabie, 1985). The ARI has the maximal value of $1$ in
the case of a perfect recovery of the underlying clustering structure,
and a value of $0$ in the case where the true membership $U$ and
estimated membership $\hat{U}$ coincide no more than would be expected
by chance. When the PO is high, the $k$-means clustering in the true
subspace defined by the true $\mathbf{A}_{1}$ does not work. Thus, in
order to calculate the ARI, the $k$-means clustering with 100 random
starts was conducted with the true $\mathbf{F}$, and then the estimated
cluster structure was considered to be the true cluster structure.

In addition, to evaluate recovery of the subspace, the root-mean-squared
error criterion was calculated:
\begin{equation*}
 RMSE=\left(\frac{1}{2}\sum_{l=1}^{2}\|\boldsymbol{v}_{l}^{*}-\hat{\boldsymbol{v}}_{l}\|_{P}^{2}\right)^{\frac{1}{2}},
\end{equation*}
where $\boldsymbol{v}_{l}^{*}$ is the true weight function,
$\hat{\boldsymbol{v}}_{l}$ is the estimate of $\boldsymbol{v}_{l}^{*}$,
and $\|\cdot\|_{P}$ denotes the norm in $\mathscr{L}^{P}$ with $P$
functional variables (see, Appendix A). In this case, the true weight
function is defined as $\boldsymbol{v}_{l}^{*}=(v_{l1}^{*},\dots,
v_{lP}^{*})$ with
$v_{l1}^{*}=\mathbf{\Phi}\mathbf{H}^{-\frac{1}{2}}\boldsymbol{a}_{Hl}$
where $\boldsymbol{a}_{Hl}$ is the $l$th column of
$\mathbf{A}_{H}=(\mathbf{A}_{1}',\mathbf{O}_{L\times M_{2}})'$, and for
$p>1$, $v_{lp}^*$ is equal to a zero function. Note that the FFKM model
has indeterminacy of rotation of weight functions, as is also true of
the FPCK model (Yamamoto, 2012). Thus, the RMSE criterion for each of
the methods was calculated after the Procrustes rotation, so that the
true weight function $\boldsymbol{v}_{l}^{*}$ was considered to be the
target.

The FFKM and FPCK methods need initial values for the parameters in the
first step of the algorithms. In our limited experience, FFKM is rather
sensitive to local optima so that it needs many initial values.  Thus,
in this simulation, we used 1000 random initial values for FFKM and 100
random initial values for FPCK. For two-step FFKM, a selection of the
number $R$ of components in the first step is needed. In this
simulation, $R$ was determined in view of cumulative percentage of the
total variation (Jolliffe, 2002) in which a selected cut-off provided
90\% cumulative variation.

\subsection{Results}

\begin{figure}[!p]
 \begin{center}
	\includegraphics[width=17cm]{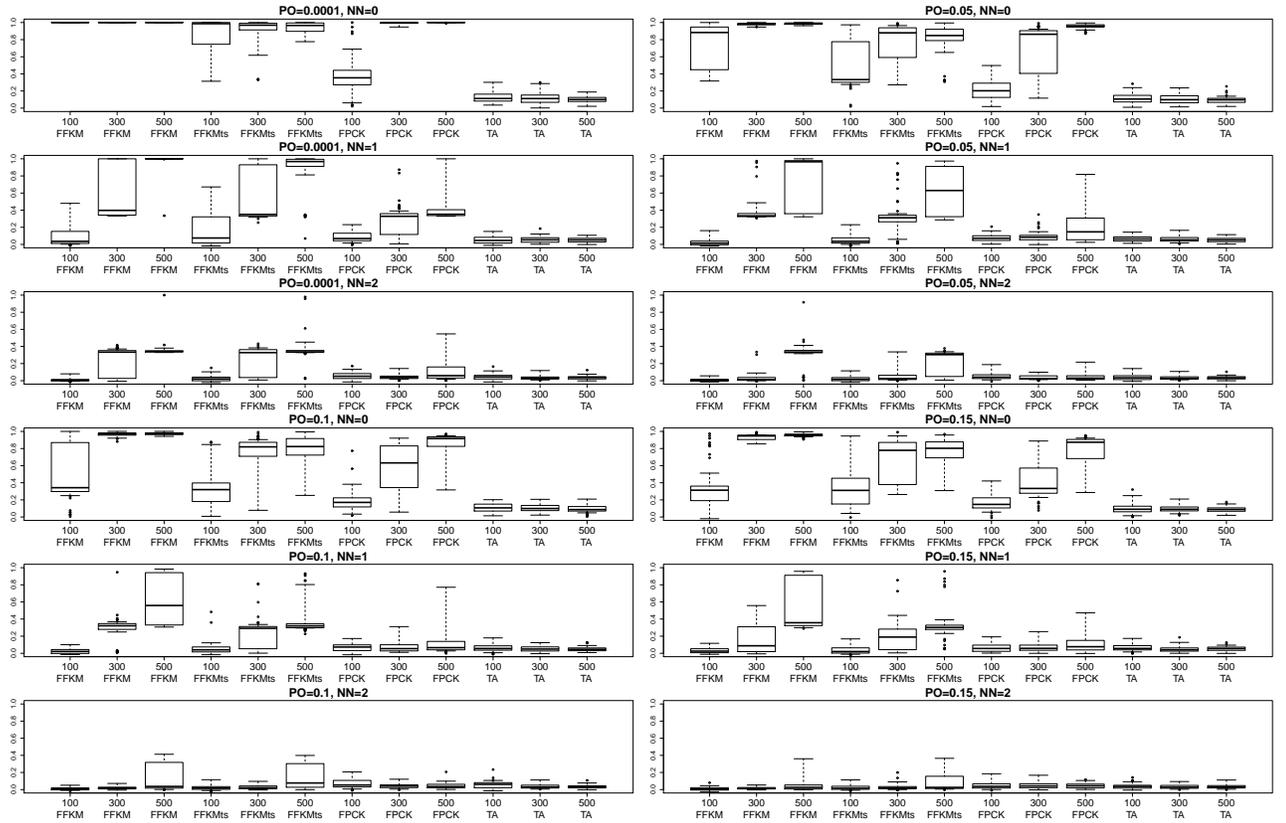}
 \caption{Boxplots of the adjusted Rand indices when $\mb{G}_{1}$ is FR
 and $\mb{G}_{2}$ is FR; in each case, from the left, the boxplots
 indicate the results of the FFKM, two-step FFKM, FPCK, and tandem
 analysis by sample size, respectively; the number above the name of
 each method in abscissa axis denotes sample size} \label{fig_ffkm:ARI1}
 \end{center}
\end{figure}

\begin{figure}[!p]
 \begin{center}
	\includegraphics[width=17cm]{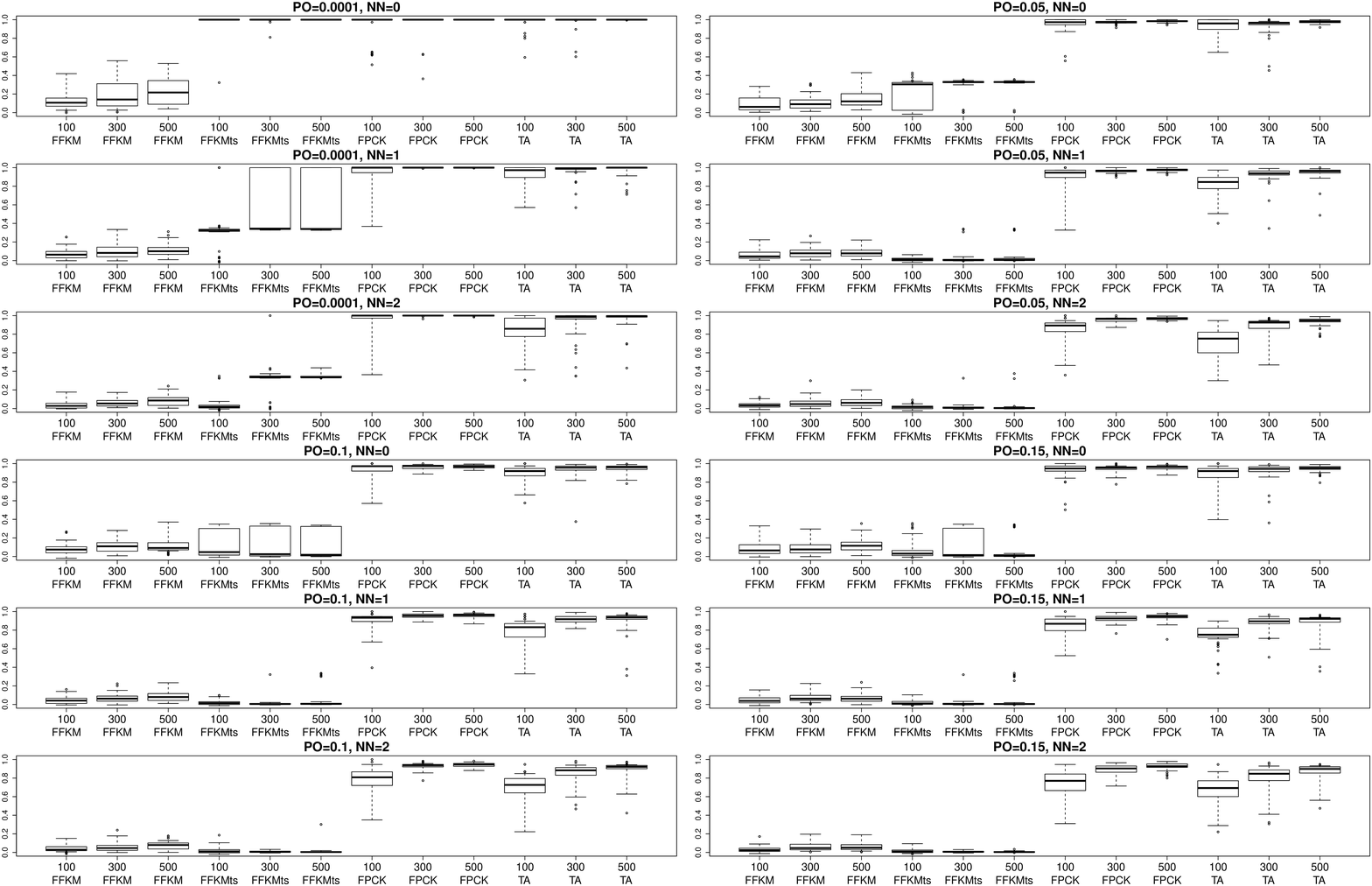}
 \caption{Boxplots of the adjusted Rand indices when $\mb{G}_{1}$ is RD
 and $\mb{G}_{2}$ is FR; in each case, from the left, the boxplots
 indicate the results of the FFKM, two-step FFKM, FPCK, and tandem
 analysis by sample size, respectively; the number above the name of
 each method in abscissa axis denotes sample size} \label{fig_ffkm:ARI2}
 \end{center}
\end{figure}

\begin{figure}[!p]
 \begin{center}
	\includegraphics[width=17cm]{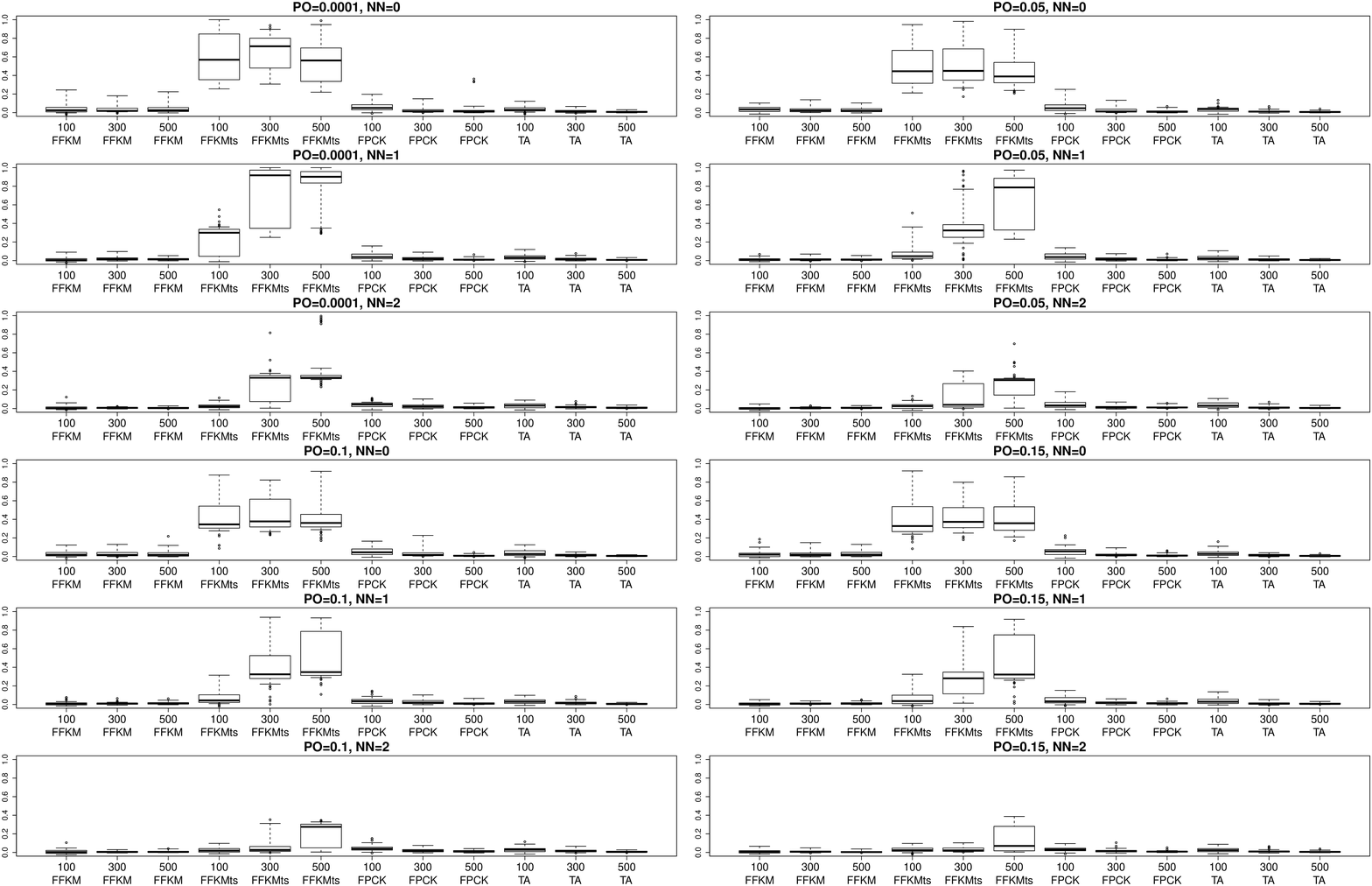}
 \caption{Boxplots of the adjusted Rand indices when $\mb{G}_{1}$ is FR
 and $\mb{G}_{2}$ is RD; in each case, from the left, the boxplots
 indicate the results of the FFKM, two-step FFKM, FPCK, and tandem
 analysis by sample size, respectively; the number above the name of
 each method in abscissa axis denotes sample size} \label{fig_ffkm:ARI3}
 \end{center}
\end{figure}

\begin{figure}[!p]
 \begin{center}
	\includegraphics[width=17cm]{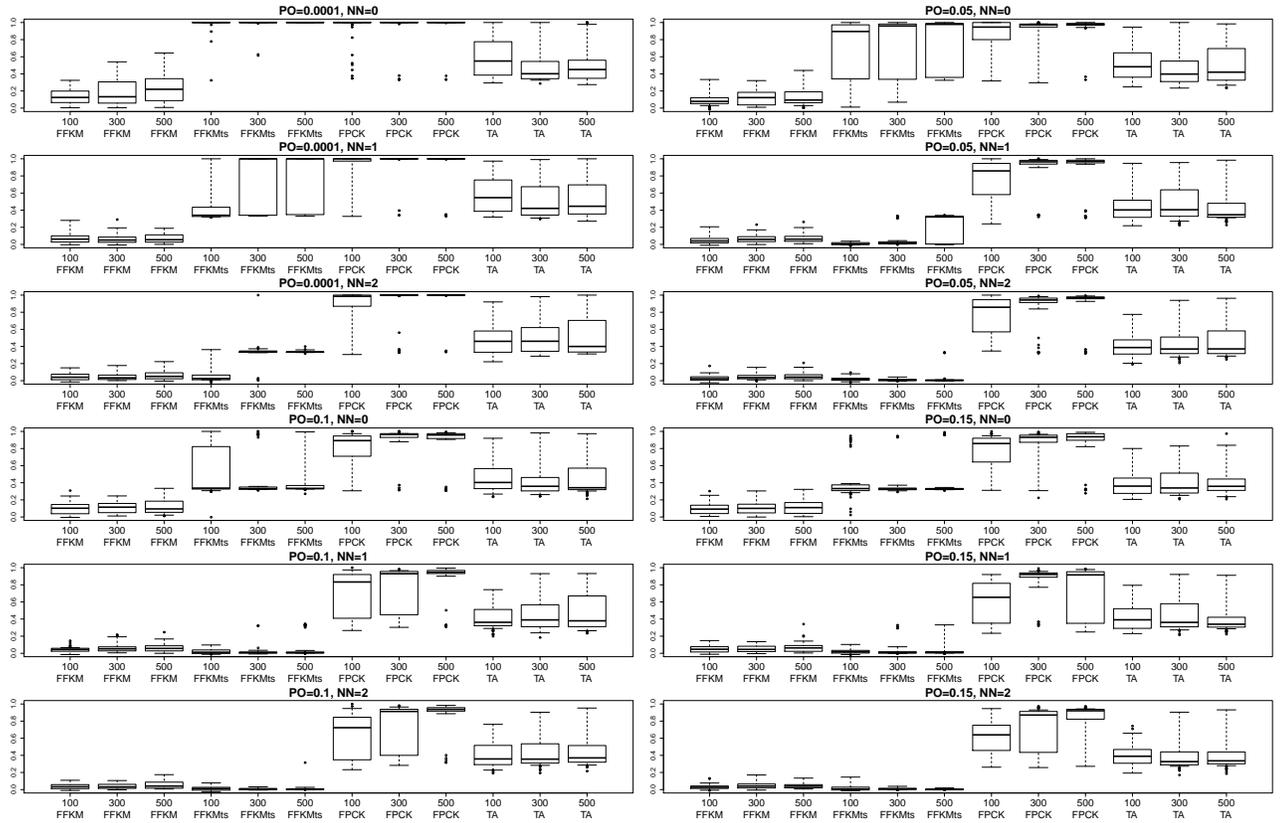}
 \caption{Boxplots of the adjusted Rand indices when $\mb{G}_{1}$ is RD
 and $\mb{G}_{2}$ is RD; in each case, from the left, the boxplots
 indicate the results of the FFKM, two-step FFKM, FPCK, and tandem
 analysis by sample size, respectively; the number above the name of
 each method in abscissa axis denotes sample size} \label{fig_ffkm:ARI4}
 \end{center}
\end{figure}

Boxplots of the ARIs obtained by the four methods are shown in Figure
\ref{fig_ffkm:ARI1}, \ref{fig_ffkm:ARI2}, \ref{fig_ffkm:ARI3}, and
\ref{fig_ffkm:ARI4} which are results for the cases of (FR, FR), (RD,
FR), (FR, RD), and (RD, RD), respectively, corresponding to the ranks of
($\mb{G}_{1}, \mb{G}_{2}$). The modified boxplot (Hubert and
Vandervieren, 2008) was used for the asymmetry of the distributions of
ARIs and RMSEs. In these figures, boxplots of four methods for each
sample size are arranged by the proportion of overlap (PO) and the
number of non-informative variables (NN). As can be inferred from Figure
\ref{fig_ffkm:ARI1}, when both $\mb{G}_{1}$ and $\mb{G}_{2}$ were FR,
under all conditions, FFKM and two-step FFKM showed the best result, or
at least a result comparable to those of the other two methods. It can
be seen that ARIs became worse with an increase in PO and NN, while the
indices improved with an increase in the sample size. FPCK also worked
well only under the easiest condition where PO was small, $N$ was large,
and there was no non-informative variable. This result shows that the
FPCK method provided a poor result if the contributing part $\mb{G}_{1}$
to the cluster structure was FR. We also see that tandem analysis did
not work well, regardless of the chosen values of PO, NN, and $N$.

When $\mb{G}_{1}$ was RD and $\mb{G}_{2}$ was FR (Figure
\ref{fig_ffkm:ARI2}), we can see that FPCK showed the best result under
all values of PO and NN, while FFKM did not. The two-step FFKM method
provided better results when $\text{PO}=0.0001$ than when PO was large,
and the ARI became worse with an increase in PO and NN. Since
$\mb{G}_{2}$ was FR and all columns of $\mb{G}_{1}$ contributed to the
cluster structure, the optimal subspace obtained from functional
principal component analysis are coincident with that obtained from
FPCK. This fact explains that tandem analysis worked as well as FPCK in
this case.

When $\mb{G}_{1}$ was FR and $\mb{G}_{2}$ was RD (Figure
\ref{fig_ffkm:ARI3}), only two-step FFKM recovered the true cluster
structure. It can be inferred that FFKM were effected negatively by the
correlation of $\mb{G}_{2}$, while two-step FFKM improved the
performance of FFKM to remove the negative effect of the cumbersome
correlation as it had been expected.

When both $\mb{G}_{1}$ and $\mb{G}_{2}$ were RD (Figure
\ref{fig_ffkm:ARI4}), FPCK showed the best result, or at least a result
comparable to those of other methods. Two-step FFKM also worked well
under mild conditions in which both PO and NN were small. FFKM and
tandem analysis did not recovered the cluster structure well because of
the existence of substantial correlation of $\mb{G}_{2}$.

\begin{figure}[!p]
 \begin{center}
	\includegraphics[width=17cm]{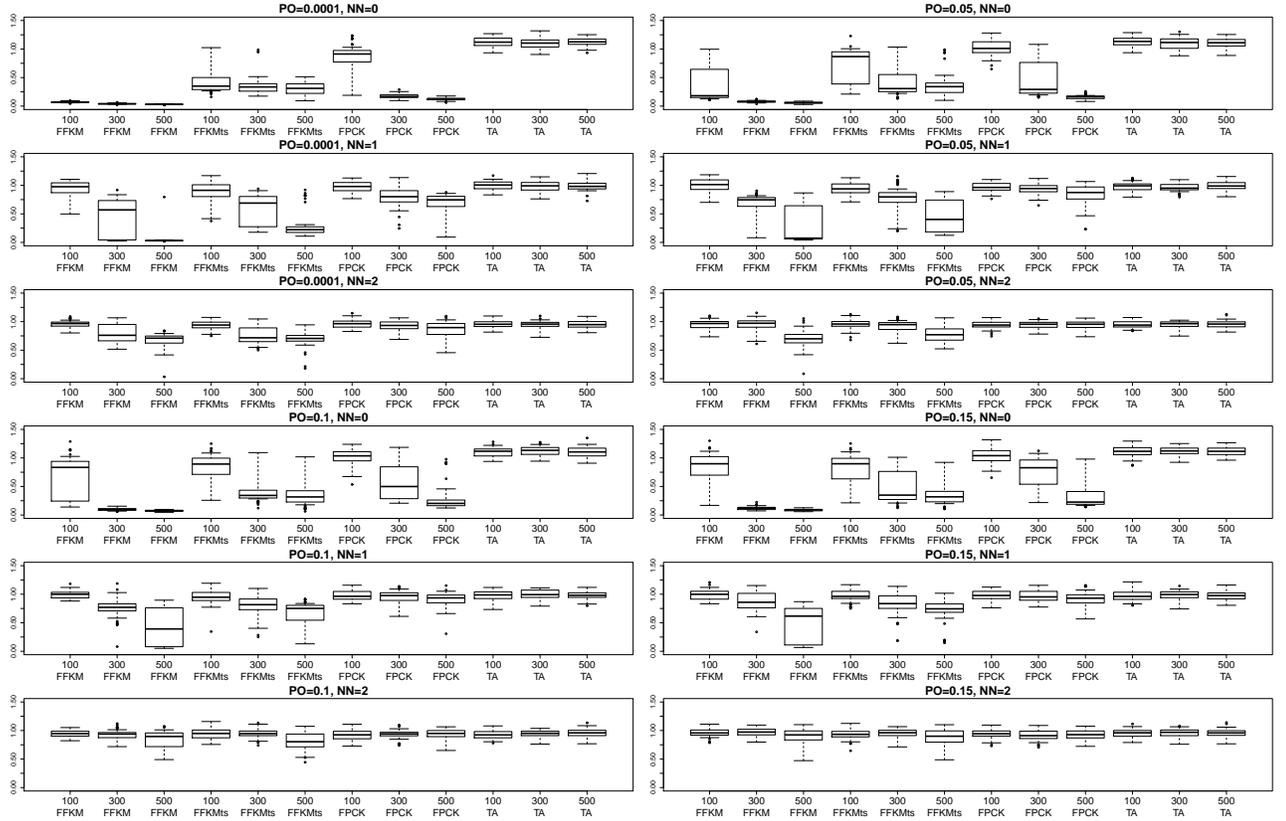}
 \caption{Boxplots of the RMSE when $\mb{G}_{1}$ is FR and $\mb{G}_{2}$
 is FR in each case, from the left, the boxplots indicate the results of
 the FFKM, two-step FFKM, FPCK, and tandem analysis by sample size,
 respectively; the number above the name of each method in abscissa axis
 denotes sample size} \label{fig_ffkm:RMSE1}
 \end{center}
\end{figure}

\begin{figure}[!p]
 \begin{center}
	\includegraphics[width=17cm]{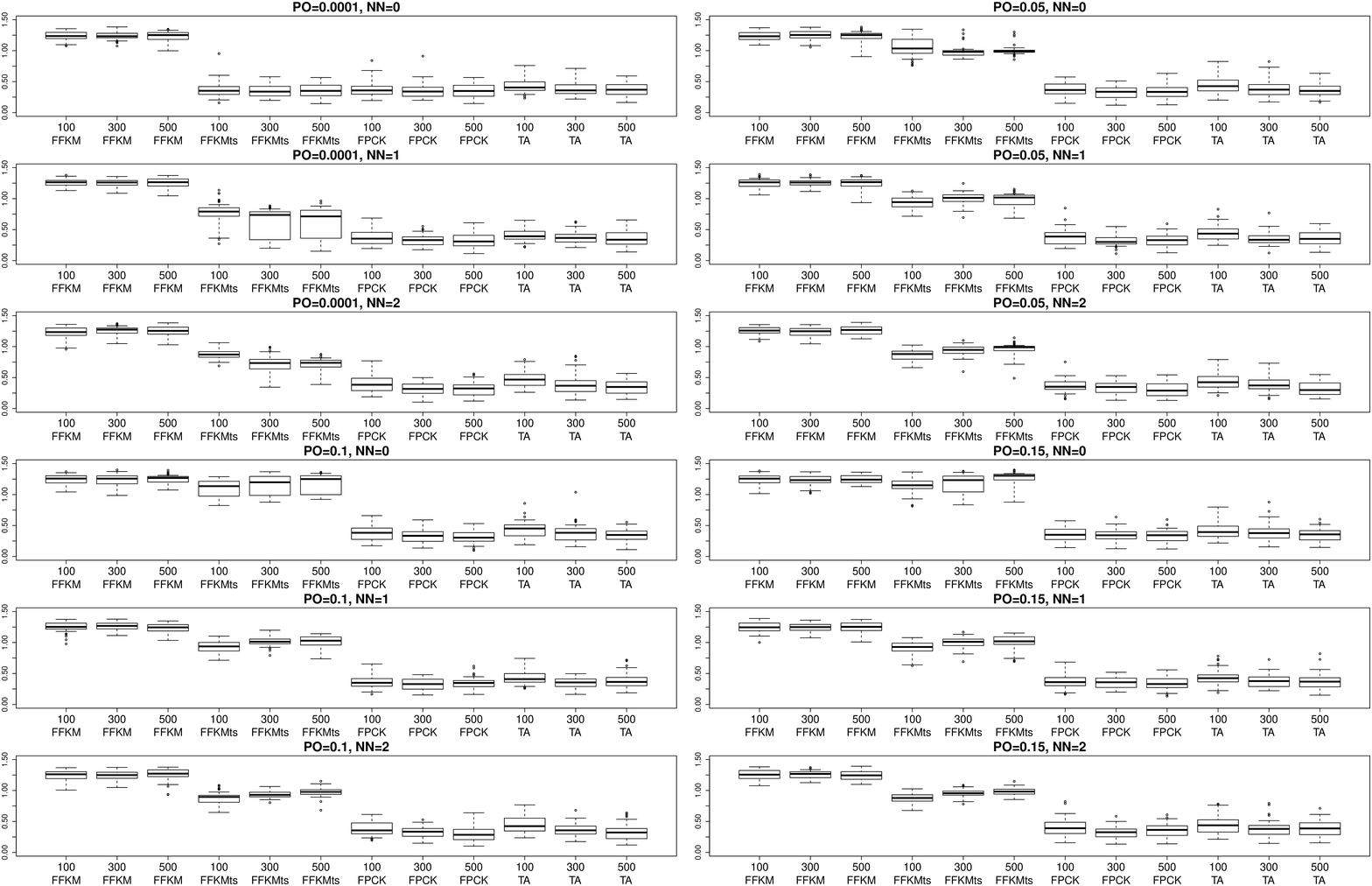}
 \caption{Boxplots of the RMSE when $\mb{G}_{1}$ is RD and $\mb{G}_{2}$
 is FR; in each case, from the left, the boxplots indicate the results
 of the FFKM, two-step FFKM, FPCK, and tandem analysis by sample size,
 respectively; the number above the name of each method in abscissa axis
 denotes sample size} \label{fig_ffkm:RMSE2}
 \end{center}
\end{figure}

\begin{figure}[!p]
 \begin{center}
	\includegraphics[width=17cm]{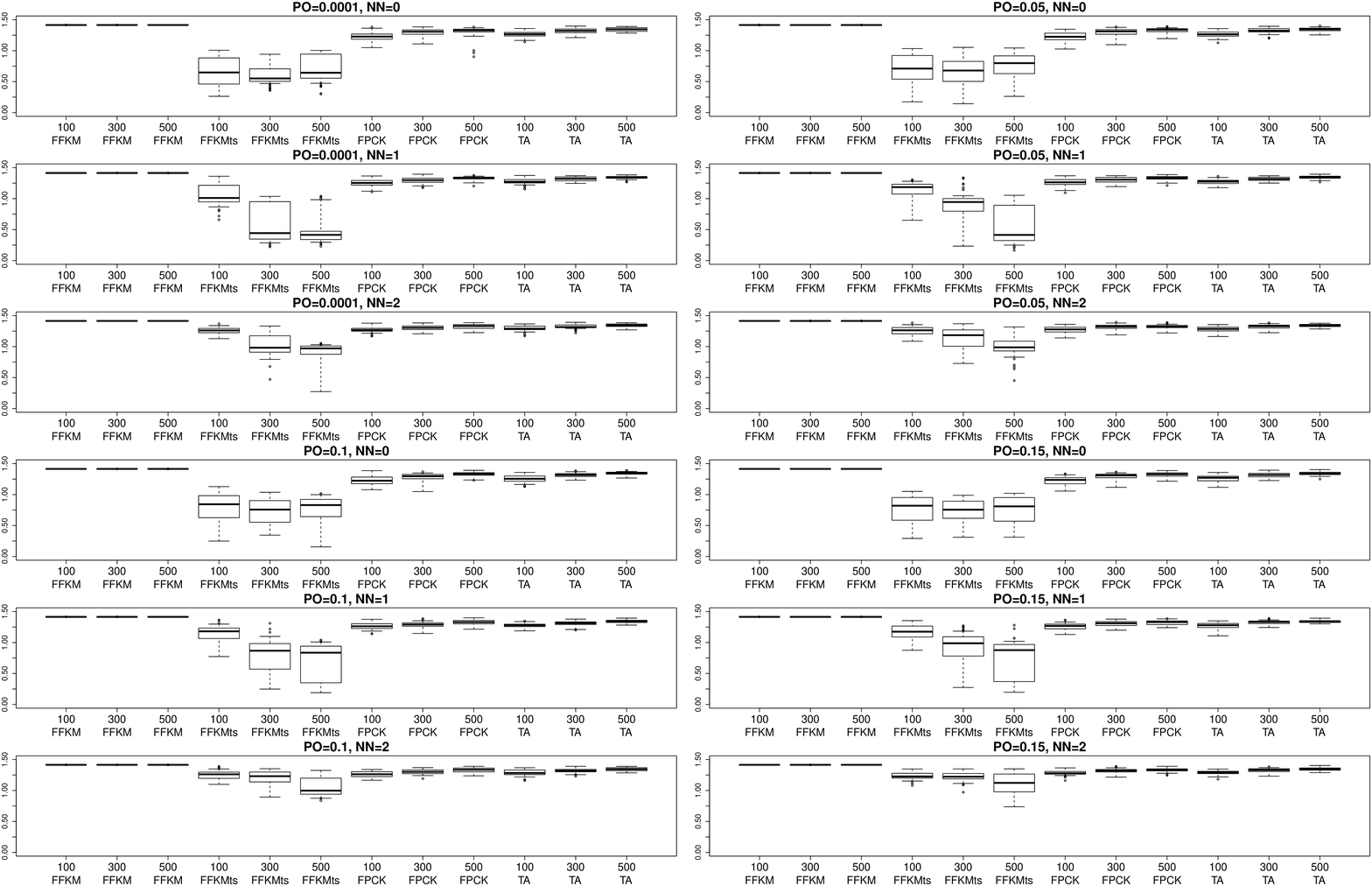}
 \caption{Boxplots of the RMSE when $\mb{G}_{1}$ is FR and $\mb{G}_{2}$
 is RD; in each case, from the left, the boxplots indicate the results
 of the FFKM, two-step FFKM, FPCK, and tandem analysis by sample size,
 respectively; the number above the name of each method in abscissa axis
 denotes sample size} \label{fig_ffkm:RMSE3}
 \end{center}
\end{figure}

\begin{figure}[!p]
 \begin{center}
	\includegraphics[width=17cm]{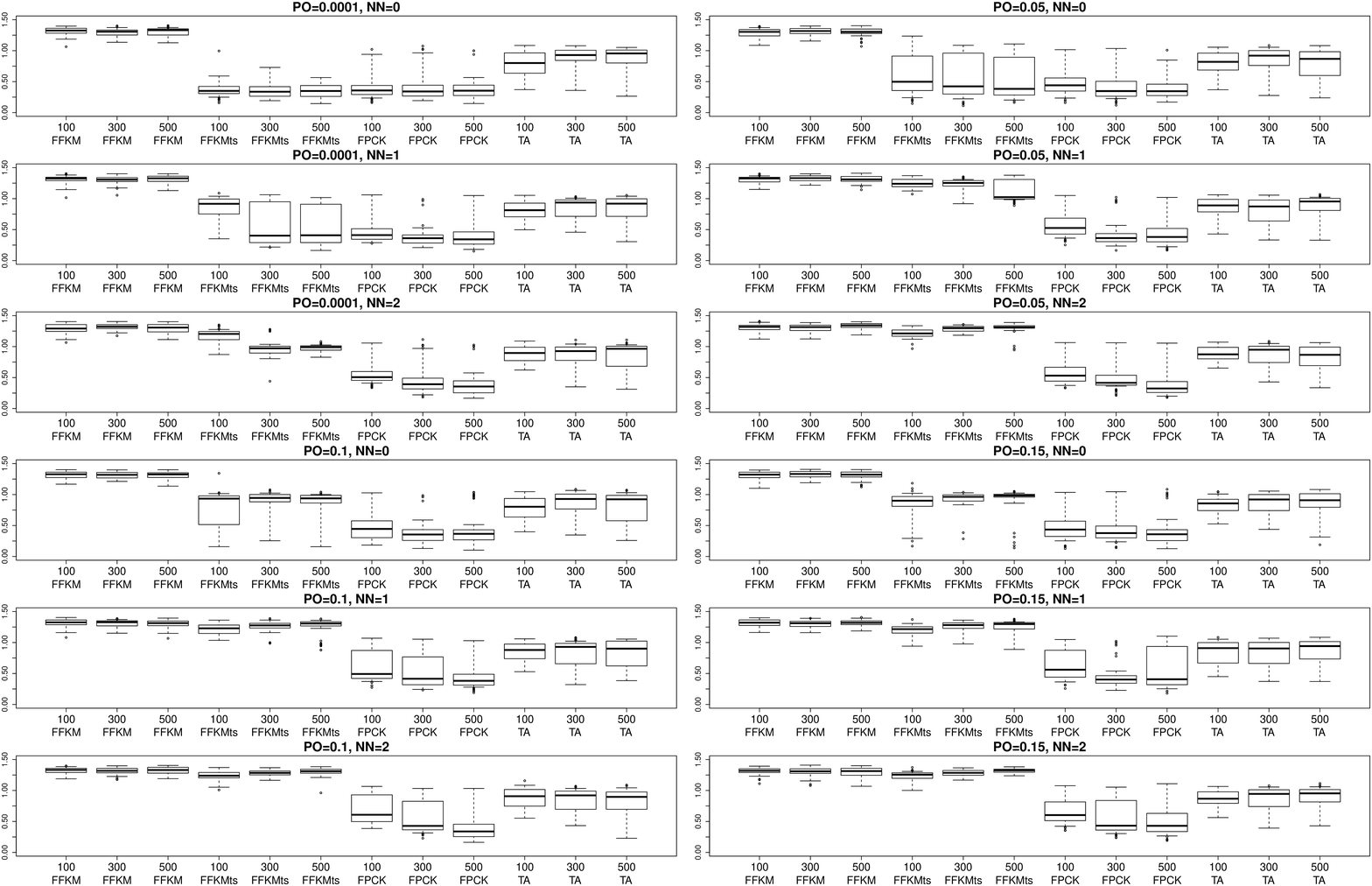}
 \caption{Boxplots of the RMSE when $\mb{G}_{1}$ is RD and $\mb{G}_{2}$
 is RD; in each case, from the left, the boxplots indicate the results
 of the FFKM, two-step FFKM, FPCK, and tandem analysis by sample size,
 respectively; the number above the name of each method in abscissa axis
 denotes sample size} \label{fig_ffkm:RMSE4}
 \end{center}
\end{figure}

Boxplots of the RMSEs obtained by the four methods are shown in Figure
\ref{fig_ffkm:RMSE1}, \ref{fig_ffkm:RMSE2}, \ref{fig_ffkm:RMSE3}, and
\ref{fig_ffkm:RMSE4} which are results for the cases of (FR, FR), (RD,
FR), (FR, RD), and (RD, RD), respectively, corresponding to the ranks of
($\mb{G}_{1}, \mb{G}_{2}$). From these figures, it can be seen that the
results of RMSEs coincide with those of ARIs.

We used 1000 random starts for the FFKM method. However, even in the
case of one of the easiest settings, where $\text{PO}=0.0001$ and
$N=500$, only 93 initial starts attained the global optimal
solution. In addition, more local optimal solutions seem to occur when
the overlap is increased. Thus, in practice, it is necessary to check
carefully whether the solution is a global optimal solution. If not,
more initial random starts may be necessary.

\section{Empirical Example}
\label{sec_ffkm:example}

In this section, we perform an empirical analysis to demonstrate the use
of the FFKM method and to compare its performance with that of the
existing methods, the FPCK and tandem analysis (TA). We used the
well-known phoneme data set for a speech-recognition problem, as
described by Hastie et al. (1995). The data are log-periodograms of 32
ms duration that correspond to five phonemes, as follows: ``sh'' as in
``she'', ``dcl'' as in ``dark'', ``iy'' as the vowel in ``she'', ``aa''
as the vowel in ``dark'', and ``ao'' as the first vowel in ``water''. We
considered only the first 150 frequencies used in Ferraty and Vieu
(2003), thus obtaining a data set of 2000 log-periodograms with the
known class-phoneme membership.

\begin{figure}[!tbp]
 \begin{center}
	\begin{tabular}{c}
	\includegraphics[width=7cm]{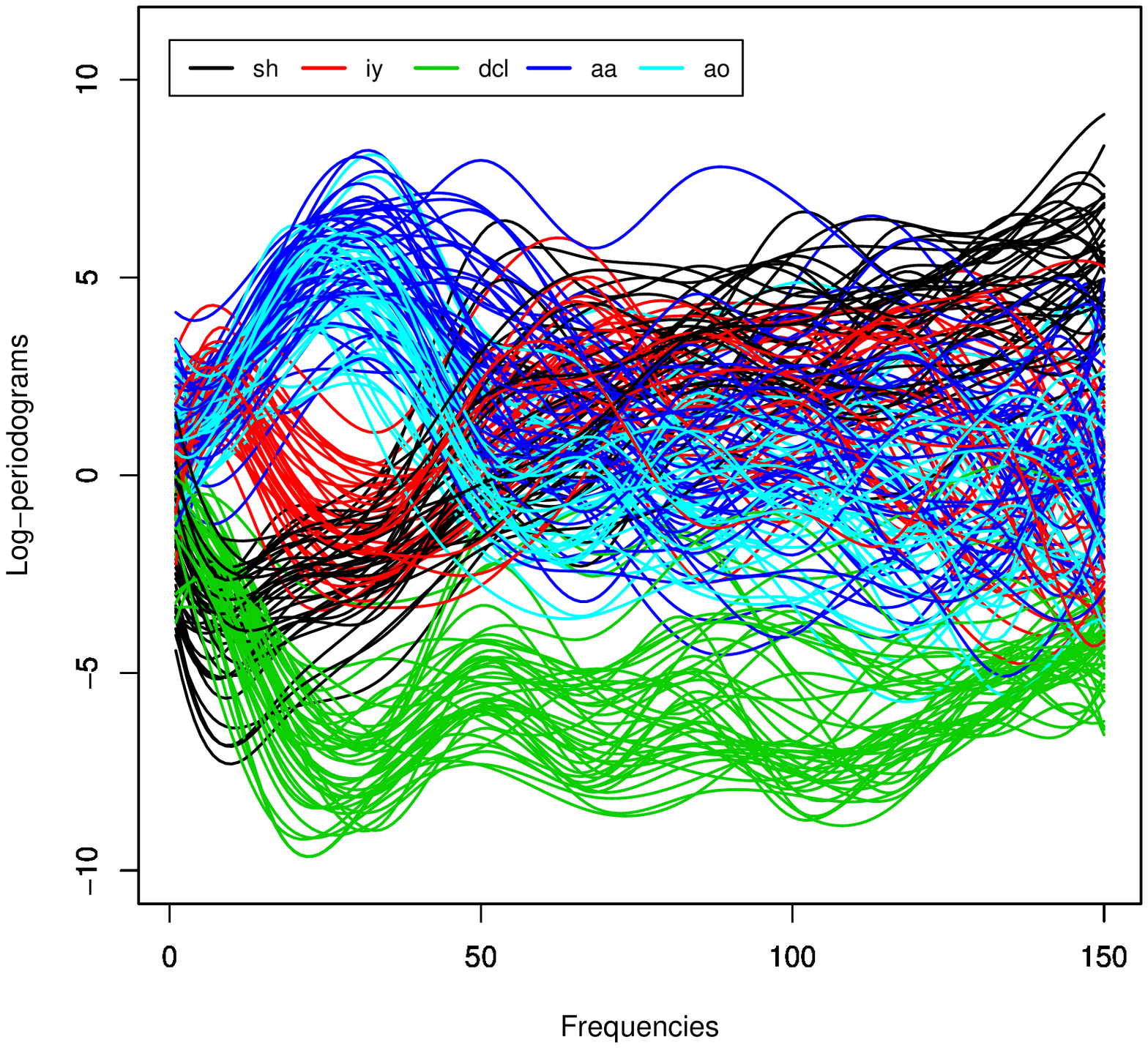}
	\end{tabular}
 \caption{Selected phoneme data of 200 log-periodograms; the color
	denotes groups of phoneme} \label{fig_ffkm:selected_phoneme_data}
 \end{center}
\end{figure}

In this example, suppose that we want to find correct clusters with $K =
5$ and obtain a low-dimensional subspace with $L = 2$ for interpreting
the cluster structure. For all methods, we used the fourth-order
B-spline basis function with ten knots. In this case, the number of
basis functions is twelve. The value of $\lambda$ that gives the minimum
of GCV among the different values of $\lambda$, varying from 0.1 to 500,
was selected: $\lambda=61.31$. The selected log-periodograms expanded by
these basis functions are shown in Figure
\ref{fig_ffkm:selected_phoneme_data}. For the FFKM and FPCK method, the
initial random starts with 100 were used.

In general, the coefficient matrix $\mathbf{G}_{H}$ of the functional
data has some correlations between the coefficient vectors corresponding
to the discretized basis functions $\phi_{m}(t)$. In such a case, there
often exist small eigenvalues, which may be nearly zero, of
$\mathbf{G}_{H}'\mathbf{G}_{H}$, so that the FFKM is likely to provide a
poor recovery of the true cluster structure. Thus, we used a two-step
approach which was investigated in Section 4.

\begin{figure}[!tb]
 \begin{center}
	\begin{tabular}{cc}
	 \includegraphics[width=5cm]{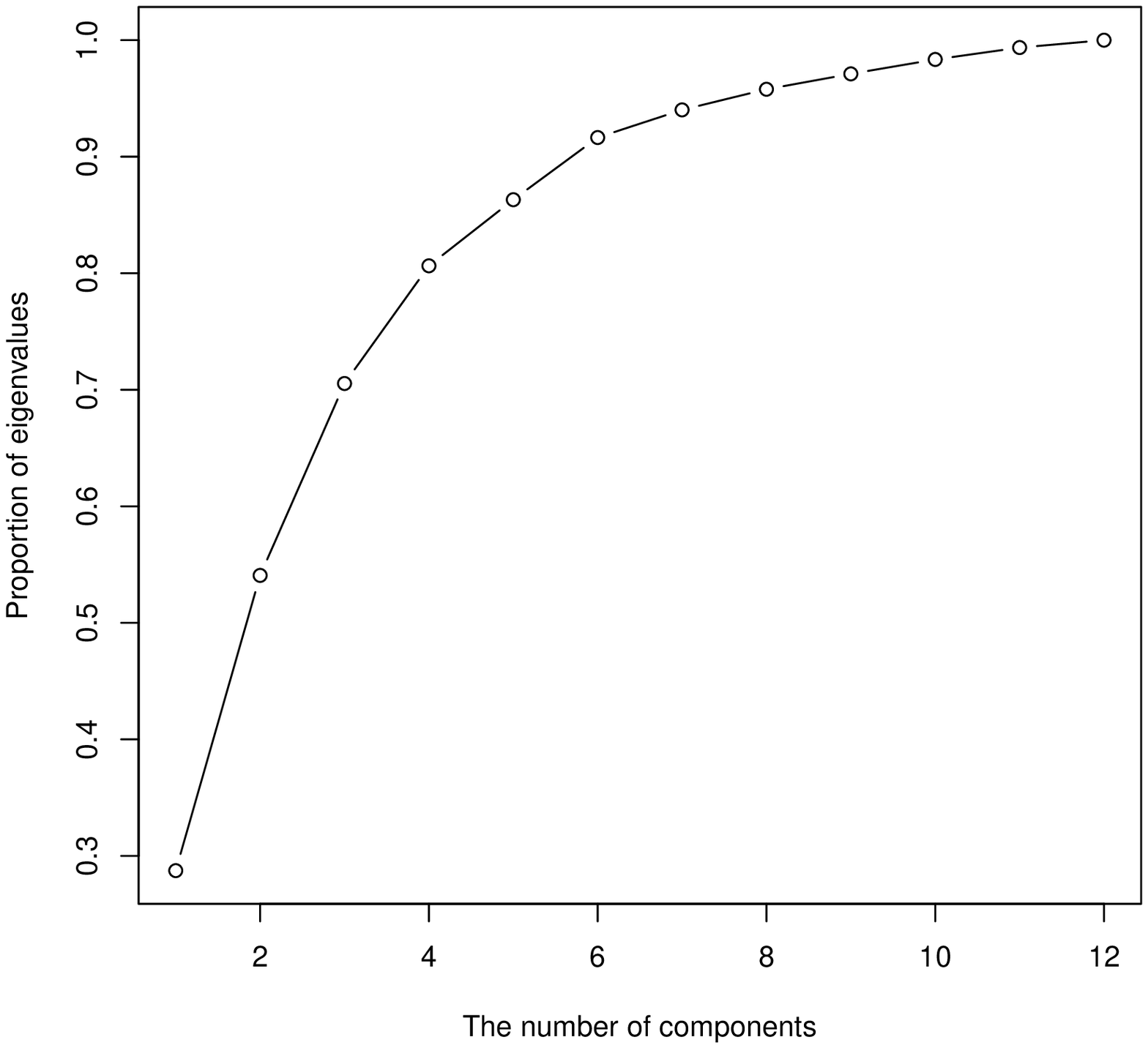} &
	 \includegraphics[width=5cm]{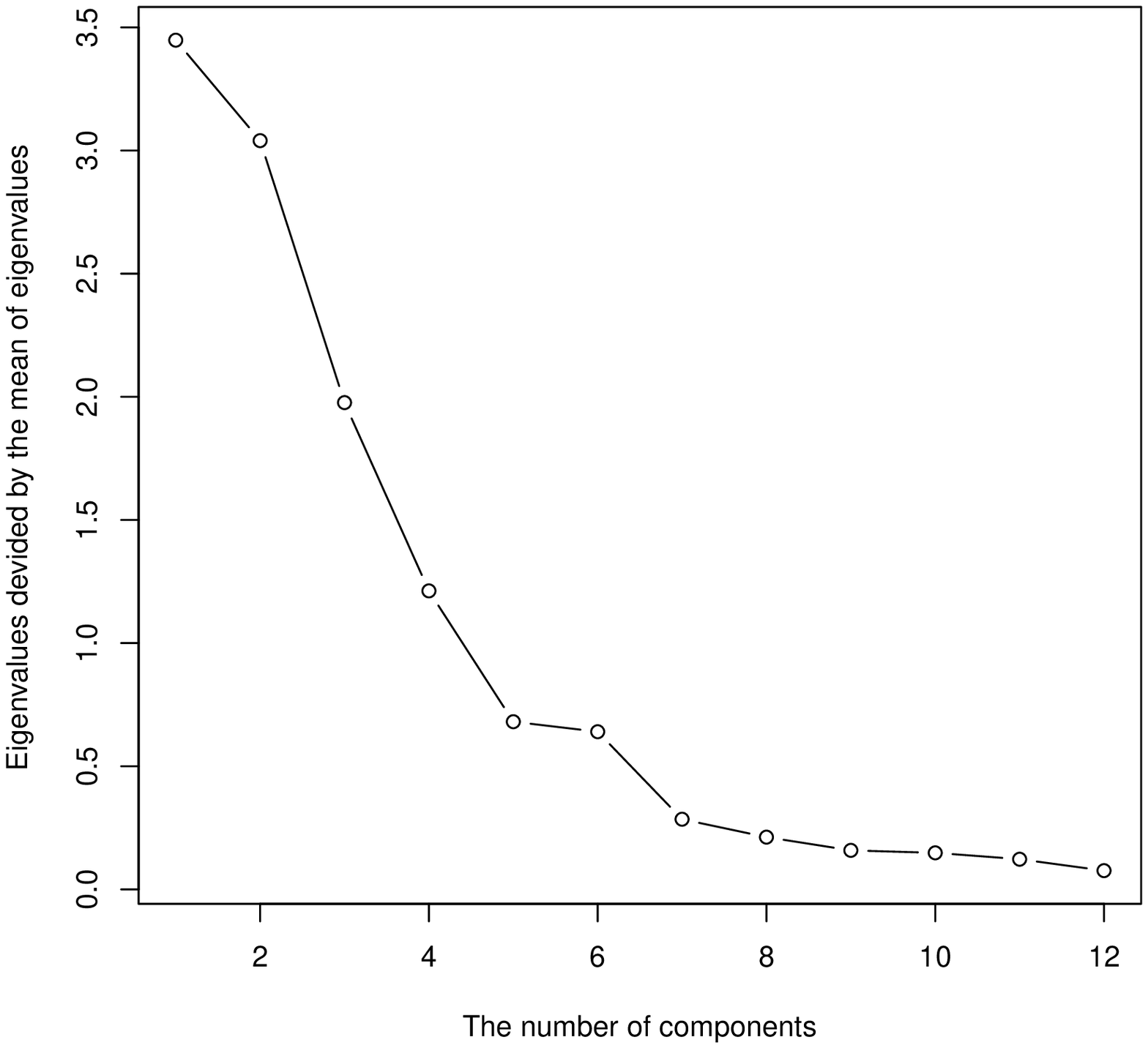}
	\end{tabular}
 \caption{Plots for justification of the number of components; the left
	denotes the proportion of eigenvalues; the right denotes the eigenvalues
	divided by their mean value}
	\label{fig_ffkm:eigenvalues}
 \end{center}
\end{figure}

In this data set, we can see that there are substantial correlations
between the columns of $\mathbf{G}_{H}$, and the simple FFKM method
provides a poor result. Thus, first we conducted FPCA with four
components; the number of components was determined by the cumulative
percentage of the total variation and the size of the variances of the
principal components, as introduced in Jolliffe (2002). In view of
cumulative percentage of the total variation, Jolliffe (2002) notes that
choosing a cut-off somewhere between $70\%$ and $90\%$ and retaining $R$
components, where $R$ is the number determined by the cut-off, provides
a rule that preserves most of the information in the data in the first
$R$ components. This is shown in the left plot of Figure
\ref{fig_ffkm:eigenvalues}. Furthermore, in view of the size of the
variances of the principal components, it is recommended that we take as
a cut-off the average value of the eigenvalues. The proportions of the
eigenvalues to the eigenvalues divided by their mean are shown in the
right plot of Figure \ref{fig_ffkm:eigenvalues}. From these plots, we
see that the chosen number, four, is justified. We therefore conducted
the FFKM analysis using the first four component scores.

The ARIs obtained by the three methods are shown in Table
\ref{tab_ffkm:ARI_phoneme}. We can see that the FFKM method can recover
the true phoneme clusters well, while the other two methods provide
cruder recoveries of the true cluster structure.
\begin{table}[!tb]
\begin{center}
 \caption{Adjusted Rand indices of the three methods}
 \label{tab_ffkm:ARI_phoneme}
\begin{tabular}{cccc} \hline
 & FFKM & FPCK & TA \\\hline
ARI & 0.599 & 0.293 & 0.293 \\\hline
\end{tabular}
\end{center}
\end{table}

\begin{figure}[!tbp]
 \begin{center}
	\begin{tabular}{ccc}
	 	FFKM & FPCK & TA \\
	 \includegraphics[width=4.5cm]{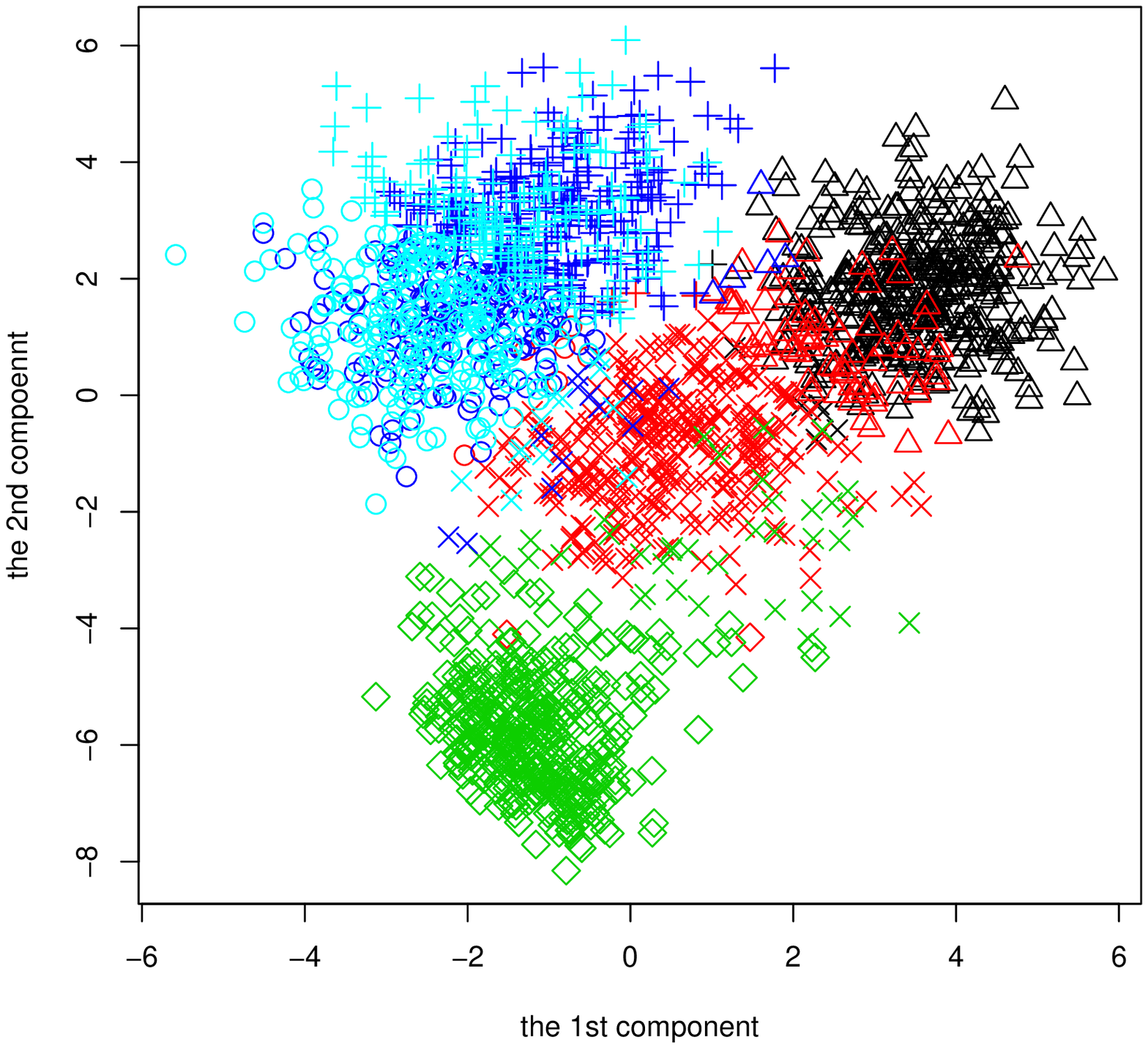} &
			 \includegraphics[width=4.5cm]{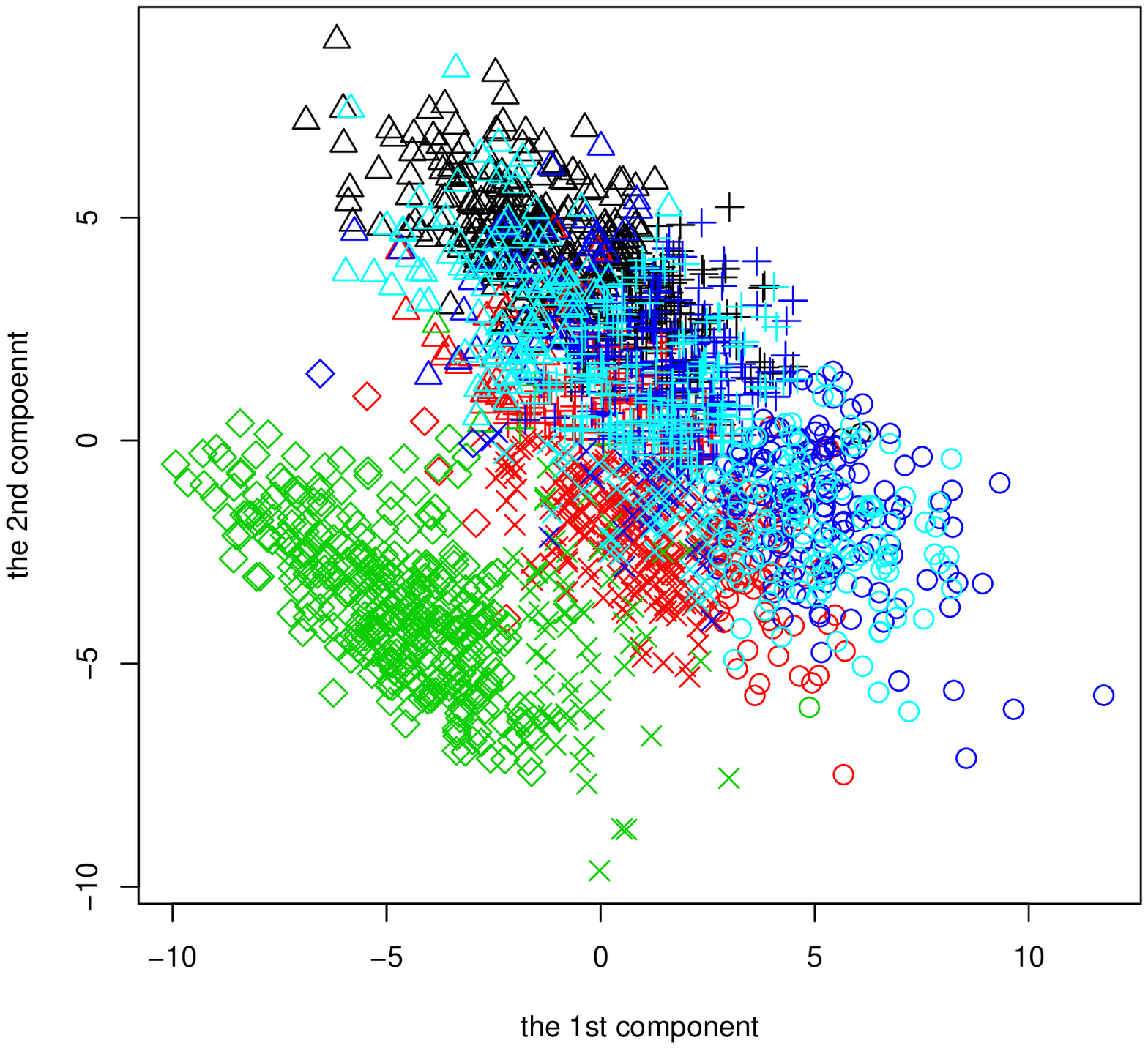} &
					 \includegraphics[width=4.5cm]{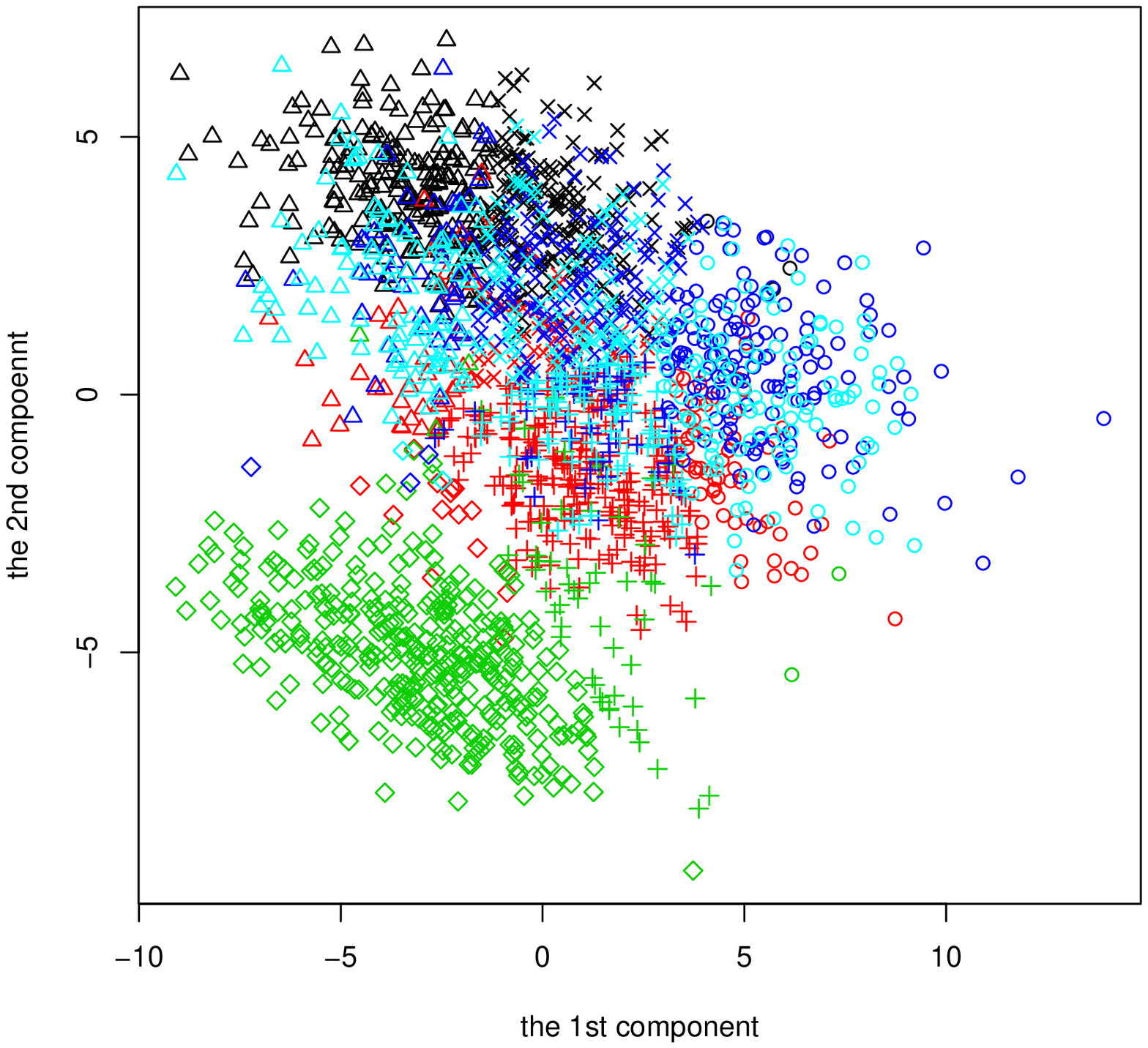}
	\end{tabular}
 \caption{Plots of the component scores estimated by the FFKM, FPCK, and
	tandem analysis; symbols of plots denote the estimated clusters of
	objects, and the colors denote the true cluster structure}
	\label{fig_ffkm:estimate_F}
 \end{center}
\end{figure}

The estimated component scores with the estimated cluster labels are
plotted in Figure \ref{fig_ffkm:estimate_F}. In each plot, the symbol
denotes the estimated clusters of objects and the colors denote the true
cluster structure. From these plots, it is concluded that the FFKM gives
the optimal subspace representing the true cluster structure, while the
subspaces given by the FPCK method and tandem analysis may not be
appropriate for finding the cluster structure.

\begin{figure}[!tbp]
 \begin{center}
	\begin{tabular}{cc}
	\includegraphics[width=6cm]{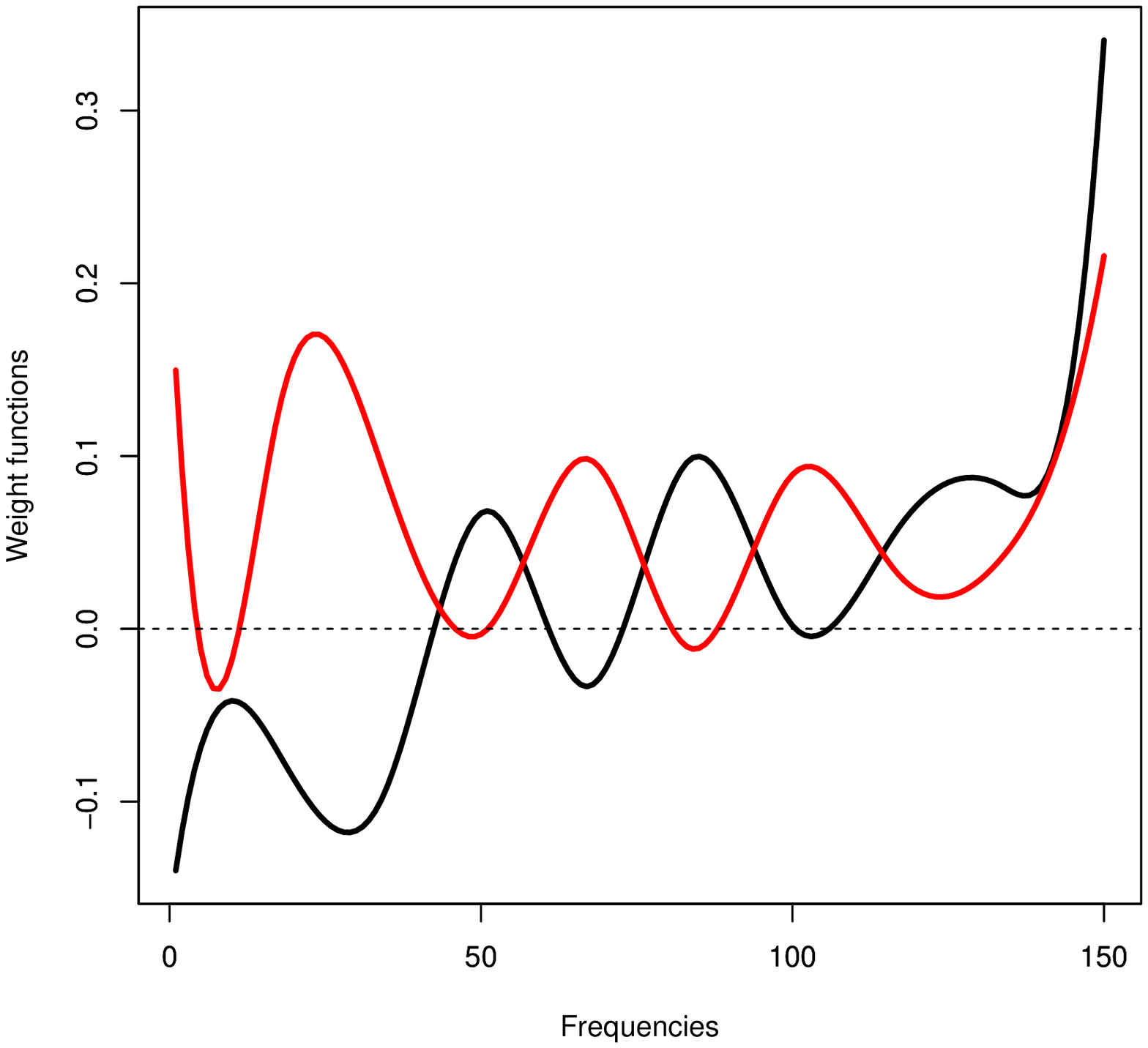} &
	 \includegraphics[width=6cm]{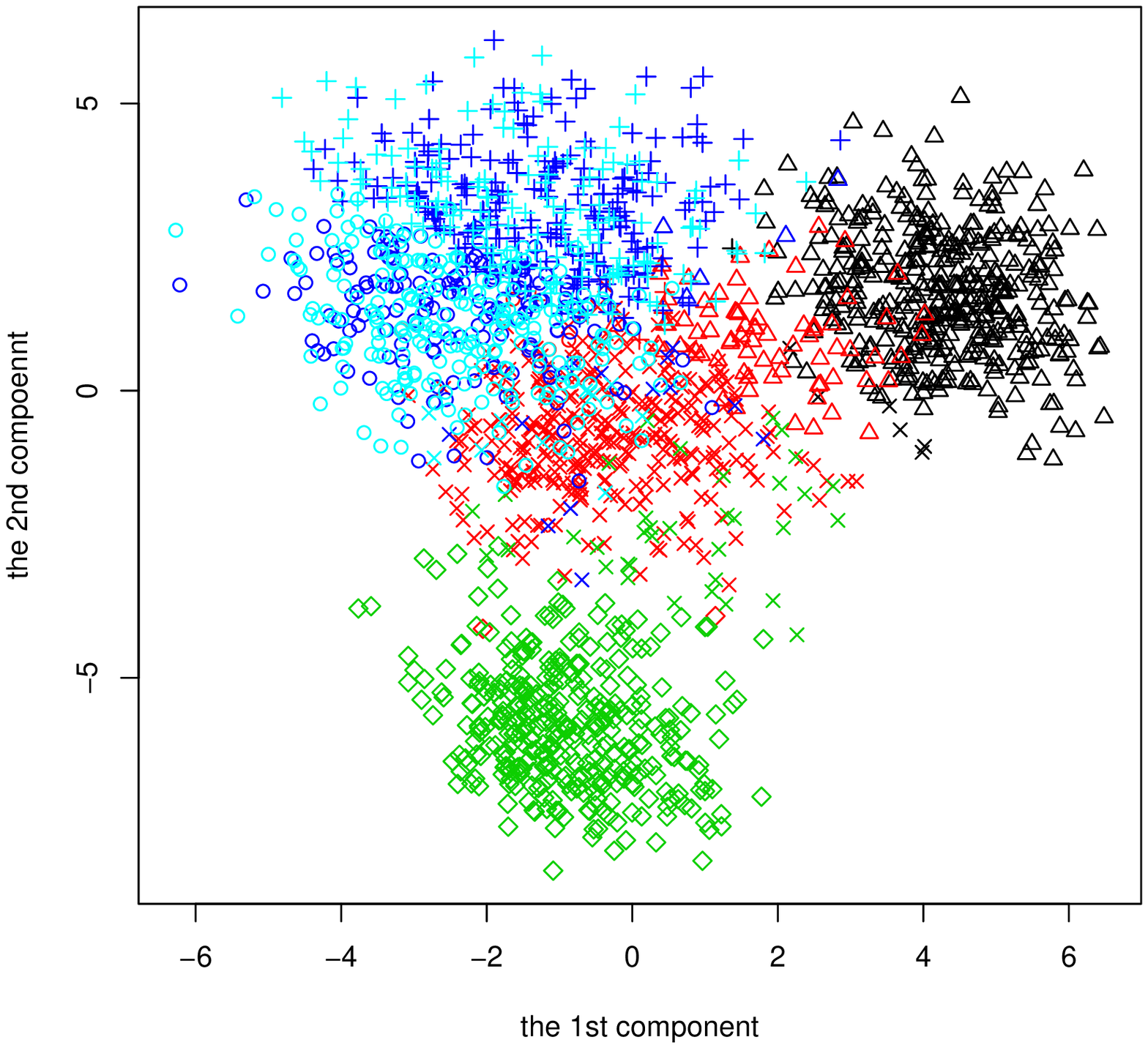}
	\end{tabular}
 \caption{Estimated weight functions (left) by the FFKM method and the
	plot of corresponding component scores (right); the black and red
	curves correspond to the first and second components,
	respectively} \label{fig_ffkm:estimate_V}
 \end{center}
\end{figure}

As with the FPCK method described by Yamamoto (2012), it may be
beneficial to interpret the estimated subspace using the estimated
weight functions $v_{l}$. The weight functions estimated by the two-step
approach are shown in Figure \ref{fig_ffkm:estimate_V}. In the figure,
the black and red curves denote the weight functions corresponding to
the first and second components, respectively. It can be seen that the
weight functions have large values in the region where the frequency is
between 10 and 50 and in the last region. This implies that the cluster
structure is determined by the behavior of the data in these regions,
and this is reasonable considering the original data that is shown in
Figure \ref{fig_ffkm:selected_phoneme_data}. Note that the component
scores shown in the right of Figure \ref{fig_ffkm:estimate_V},
calculated using these estimated weight functions, may be a little bit
different from the original subspace representation shown in Figure
\ref{fig_ffkm:estimate_F}. In this case, however, the cluster structure
seems to be the same as the original one shown in Figure
\ref{fig_ffkm:estimate_F}. This difference is due to the method of
estimating the weight functions in the two-step approach described in
Appendix B.

Note that most of the solutions of FFKM analysis given by initial random
starts attained the same values for the loss functions. Thus, in this
case, the number of initial random starts is sufficient to obtain the
global solution.

\section{Discussion}

In this article, we explained the drawbacks of the FPCK method and
proposed a new method, FFKM analysis, to overcome the problem. The FFKM
method aims to simultaneously classify functional objects into optimal
clusters and find a subspace that best describes the classification and
dimension reduction of the data. The ALS algorithm was proposed to
efficiently solve the minimization problem of the least-squares
objective function. Analyses of artificial data reveal that the FFKM
method can give an optimal cluster structure when both the coefficient
matrix, $\mb{G}_{1}$, which is related to the true cluster structure,
and a non-informative part, $\mb{G}_{2}$, have no substantial
correlation.

However, the simulation study in Section 4 showed that when either
$\mb{G}_{1}$ or $\mb{G}_{2}$ is rank deficient, FFKM failed in providing
an optimal cluster structure. To avoid the negative effect of
correlation among $\mb{G}_{H}=(\mb{G}_{1},\mb{G}_{2})$, the two-step
approach to FFKM was also described. Two-step FFKM aims to eliminate
trivial dimensions followed by applying the FFKM algorithm to the
reduced functional data. The simulation study showed that when
$\mb{G}_{1}$ was full rank and $\mb{G}_{2}$ was rank deficient, two-step
FFKM recovered a cluster structure well. Furthermore, when $\mb{G}_{1}$
was rank deficient, it worked well under the mild conditions regardless
of the rank of $\mb{G}_{2}$. Thus, in practice, it is recommended to use
two-step FFKM instead of simple FFKM.

The simulation study also showed that when $\mb{G}_{1}$ was rank
deficient, FPCK worked well regardless of the rank of
$\mb{G}_{2}$. However, it did not work very well when $\mb{G}_{1}$ was
full rank. Specifically, if $\mb{G}_{1}$ was full rank and $\mb{G}_{2}$
was rank deficient, it did not recover the true cluster structure at
all. On the other hand, in the situation, only two-step FFKM worked
well. This fact shows that FFKM has a mutually complementary
relationship with FPCK. In practical situations, $\mb{G}_{2}$ often has
a substantial correlation, that is, $\mb{G}_{2}$ is likely to be rank
deficient. Therefore, it is recommended that first the two-step FFKM
method is implemented. If the result does not seem to be good, then FPCK
is implemented.

Both the FFKM and FPCK methods need several initial random starts for
the parameters in order to avoid local optima. In our limited
experience, this problem seems to be more serious for the FFKM
method. Thus, a more efficient algorithm for this model is needed.

In our approach, the tuning of the smoothing parameter, $\lambda$, is
done by applying the GCV criterion to each curve, and the real data
example introduced in Section \ref{sec_ffkm:example} shows that this
approach works well for finding the cluster structure. Another approach
can also be adopted. For example, Gattone and Rocci (2012) proposed an
automatic smoothing algorithm in which the smoothing is carried out
within the clustering, and the amount of smoothing is determined
adaptively. Recently, Wang (2010) has proposed a method based on
clustering instability for selecting the number of clusters. These
approaches may be applicable to the selection of the model in FFKM. This
is an area of future research that we intend to pursue.

 \section*{Acknowledgment}
We thank the Associate Editor and an anonymous reviewer for their
constructive comments that helped to improve the quality of this
article. This work was supported by JSPS Grant-in-Aid for JSPS Fellows
Number 24-2676.

\appendix

\renewcommand{\theequation}{A.\arabic{equation}}
\setcounter{equation}{0}
\section*{Appendix A: FFKM for multivariate functional data}
\label{sec_ffkm:multivar}

The method for the univariate case has been described above. Here, we
explain our method for the multivariate case. Let $\mathscr{L}^{P}$ be
the Cartesian product of $P$ sets of $\mathscr{L}=L^{2}(T)$. Then,
subject $n$ has $P$ functions, $x_{n}=(x_{n1},\dots,x_{nP}) \in
\mathscr{L}^{P}$, and an inner product for $x,y\in\mathscr{L}^{P}$ is
redefined as
\begin{equation}
 \langle x,y \rangle_{P,\lambda} =
	\sum_{p=1}^{P}\left(\int_{T}x_{p}(t)y_{p}(t)dt
								+\lambda\int_{T}\mathtt{D}^{2}x_{p}(t)\mathtt{D}^{2}y_{p}(t)dt\right).
	\label{eq_ffkm:inner_product_multivariate}
\end{equation}
Note that the norm $\|\cdot\|_{P,\lambda}$ is given by the inner
product, i.e.,
$\|x\|_{P,\lambda}=\left<x,x\right>_{P,\lambda}^{1/2}$. Then, the
objective function in Eq. (\ref{eq_ffkm:target}) will be optimized as in
the univariate case, with $\mathtt{P}_{v}^{P}$ for $\mathtt{P}_{v}$,
where $\mathtt{P}_{v}^{P}$ is an orthogonal projection operator from
$\mathscr{L}^{P}$ onto the subspace $\mathscr{S}_{v}^{P}$, and the
weight functions $v_{l}^{P}\in\mathscr{L}^{P}$ span
$\mathscr{S}_{v}^{P}$. That is, the loss function for the multivariate
regularized case can be written as
\begin{equation*}
L_{ffkm}(U,V)=\sum_{n=1}^{N}\sum_{k=1}^{K}u_{nk}\|\mathtt{P}_{v}\mathtt{S}^{2}_{\lambda}x_{n}-\mathtt{P}_{v}\mathtt{S}_{\lambda}^{2}\bar{x}_{k}\|_{P,\lambda}^{2}.
\end{equation*}
Here, we can consider the basis function expansion for
$\mathtt{S}_{\lambda}^{2}x_{n}$ as
\begin{align}
 \mathtt{S}_{\lambda}^{2}x_{n}=(\bs{g}_{n1}'\bs{\phi},\dots,
 \bs{g}_{nP}'\bs{\phi}),
\end{align}
where $\bs{g}_{np}$ is a coefficient vector for the basis function
expansion of $\mathtt{S}_{\lambda}^{2}x_{np}$.
Then, the criterion $L_{ffkm}$ can be derived by
\begin{align*}
 L_{ffkm}(U,V)&=\sum_{n=1}^{N}\sum_{k=1}^{K}u_{nk}\sum_{p=1}^{P}\|(\mathtt{P}_{v}\mathtt{S}_{\lambda}^{2}x_{n})-(\mathtt{P}_{v}\mathtt{S}_{\lambda}^{2}\bar{x}_{k})_{p}\|_{P,\lambda}^{2}\\
 &=\sum_{n=1}^{N}\sum_{k=1}^{K}u_{nk}\sum_{p=1}^{P}\left\{\sum_{l=1}^{L}\left(\sum_{p'=1}^{P}\boldsymbol{a}_{lp'}'\mathbf{H}_{\lambda}\boldsymbol{g}_{np'}\right)\boldsymbol{a}_{lp}-\sum_{l=1}^{L}\left(\sum_{p'=1}^{P}\boldsymbol{a}_{lp'}'\mathbf{H}_{\lambda}\bar{\boldsymbol{g}}_{kp'}\right)\boldsymbol{a}_{lp}\right\}'\\
 &\ \ \ \ \ \mathbf{H}_{\lambda}\left\{\sum_{l=1}^{L}\left(\sum_{p'=1}^{P}\boldsymbol{a}_{lp'}'\mathbf{H}_{\lambda}\boldsymbol{g}_{np'}\right)\boldsymbol{a}_{lp}-\sum_{l=1}^{L}\left(\sum_{p'=1}^{P}\boldsymbol{a}_{lp'}'\mathbf{H}_{\lambda}\bar{\boldsymbol{g}}_{kp'}\right)\boldsymbol{a}_{lp}\right\}.
\end{align*}

The algorithm to minimize the objective function for multivariate
functional data is the same as that for univariate functional data
described above, i.e., the ALS algorithm can be applied, although there
are some differences between these cases. In {\it STEP2}, the basis
function expansion of a projected object
$\mathtt{P}_{v}^{P}\mathtt{S}_{\lambda}x_{n}$ can be applied as in the
case of univariate data. Thus, the cluster parameters $U$ are estimated
using the $k$-means algorithm for a parameter vector
$\boldsymbol{d}^{*}=(\boldsymbol{d}_{1}',\dots,\boldsymbol{d}_{N}')'$,
where $\boldsymbol{d}_{np}=(d_{np1},\dots,d_{npM})'$ and
$\boldsymbol{d}_{n}=(\boldsymbol{d}_{n1}',\dots,\boldsymbol{d}_{nP}')'$,
which is the parameter vector for the basis function expansion of
$\mathtt{P}_{v}^{P}\mathtt{S}_{\lambda}x_{n}$.

Next, we consider the optimization over $V$. Let an integral operator
$\mathtt{F}^{P}$ be defined, for any $y\in\mathscr{L}^{P}$, as
\begin{equation*}
 \mathtt{F}^{P}y:=(\mathtt{F}^{(1)}y,\dots,\mathtt{F}^{(P)}y),
\end{equation*}
where
\begin{equation*}
	(\mathtt{F}^{(p)}y)(t):=-\sum_{n=1}^{N}\sum_{k=1}^{K}u_{nk}(x_{np}(t)-\bar{x}_{kp}(t))
	\sum_{p'=1}^{P}\langle x_{np'}-\bar{x}_{kp'},y_{p'}\rangle\hspace{10pt}(p=1,\dots,P).
\end{equation*}
Then, to estimate an optimal $V$ in {\it STEP3} of the above ALS
algorithm, the following optimization problem is considered:
\begin{align*}
 \max_{V}\sum_{l=1}^{L}\langle v_{l},\mathtt{F}^{P}v_{l}\rangle_{P,\lambda}.
\end{align*}
As with the univariate case, it can be verified that the operator
$\mathtt{F}^{P}$ is self-adjoint and compact. Thus, optimizing the
criterion is equivalent to solving the following eigenvalue equation,
\begin{equation*}
 \mathtt{F}^{P}\mathbf{\xi}_{l}=\rho_{l}\mathbf{\xi}_{l},\hspace{10pt}\text{
	subject to}\vspace{10pt}\left<\xi_{l},\xi_{l'}\right>_{P,\lambda}=\delta_{ll'}.
\end{equation*}

Let $\mathbf{G}_{p}=(\boldsymbol{g}_{1p},\dots,\boldsymbol{g}_{Np})'$
and
$\mathbf{G}_{Hp}=\mathbf{G}_{p}\mathbf{H}_{\lambda}^{\frac{1}{2}}$. Let
$\mathbf{G}_{H}^{P}$ be the block diagonal matrix that has
$\mathbf{G}_{Hj}$ for the $j$th diagonal block, and let
$\boldsymbol{a}_{Hl}^{P}=(\boldsymbol{a}_{Hl1}',\dots,\boldsymbol{a}_{HlP}')'$. Then,
the above eigenvalue equation reduces to
\begin{equation*}
 \mathbf{G}_{H}^{P}{}'(\boldsymbol{1}_{P}\boldsymbol{1}_{P}'\otimes(\mathbf{P}_{U}-\mathbf{I}_{N}))
	\mathbf{G}_{H}^{P}\boldsymbol{a}_{Hl}^{P}=\rho\boldsymbol{a}_{Hl}^{P}.
\end{equation*}
Finally, the estimated weight function can be calculated as
$v_{lp}=\boldsymbol{a}_{Hlp}'\mathbf{H}^{-\frac{1}{2}}\boldsymbol{\phi}$.

\renewcommand{\theequation}{B.\arabic{equation}}
\setcounter{equation}{0}
\section*{Appendix B: Two-step approach for FFKM}
\label{sec_ffkm:appendix2}

In practice, the coefficient matrix $\mathbf{G}_{H}$ of the functional
data often has high correlations between the coefficient vectors
corresponding to the discretized basis functions $\phi_{m}(t)$. In such
a case, there often exist small eigenvalues, which may be zero or nearly
zero, of $\mathbf{G}_{H}'\mathbf{G}_{H}$, so that the FFKM is likely to
adopt the eigenspace corresponding to the small eigenvalues. This can be
confirmed by the inequality (\ref{eq_ffkm:inequality}) in which the
left-hand side is the loss function of FFKM and the right-hand side is
the sum of the squared norm of projected functional data. The right-hand
side is equivalent to the sum of variances of object scores. From this
inequality, it can be seen that when an empirical covariance operator of
functional data has excessively small eigenvalues compared with the
others, the subspace spanned by eigenfunctions corresponding to the
small eigenvalues provides the smallest value of loss function of
FFKM. This results in poor recovery of the true cluster
structure. Actually, this problem also occurs in the factorial $k$-means
(Vichi and Kiers, 2001) for a usual data matrix, and it is recommended
that such trivial dimensions could be first eliminated from the
data. Thus, it is inferred that the direct use of the FFKM method may
fail to find an optimal cluster structure. To overcome this problem, we
propose the two-step approach described below.  Note that this two-step
approach has a completely different aim from that of tandem analysis:
tandem analysis finds a low-dimensional subspace regardless of the
cluster structure, whereas the two-step approach just eliminates the
trivial dimensions and finds a low-dimensional subspace where a cluster
structure exists.

First, we conduct FPCA (Ramsay and Silverman, 2005) based on the basis
function expansion using the basis function $\{\phi_{m}\}_{m=1,\dots,M}$
of the raw data.  This gives the principal curves
$\{w_{r}\}_{r=1,\dots,R}$, where $R$ is the number of principal
components. The number $R$ should be selected so that principal
components contain sufficiently high variances for the data. The usual
selection rules described by Jolliffe (2002) may work well. Let
$\mathtt{P}_{w}$ be the operator that projects functional objects onto
the space spanned by the principal curves, and we then obtain the
projected functional data $\mathtt{P}_{w}x_{i}$. As described in
Eq. (\ref{eq_ffkm:expansion_data}), the FFKM method requires a basis
function expansion of the data. Here, using the basis functions
$\phi_{m}$ used in the FPCA, the reduced functional data can be
expressed as
\begin{equation}
 (\mathtt{P}_{w}x_{1},\dots,\mathtt{P}_{w}x_{N})'
	=\mathbf{G}_{H}\mathbf{B}_{H}\mathbf{B}_{H}'\mathbf{H}^{-\frac{1}{2}}\boldsymbol{\phi},
\end{equation}
where $\mathbf{B}_{H}=(\boldsymbol{b}_{H1},\dots,\boldsymbol{b}_{HR})$
denotes the coefficient matrix in the basis function expansion of the
principal curves, such that
$w_{r}=\boldsymbol{b}_{Hr}'\mathbf{H}^{-1/2}\boldsymbol{\phi}$. In this
notation, the principal component score matrix is calculated as
$\mathbf{F}_{pca}=\mathbf{G}_{H}\mathbf{B}_{H}$.

Thus, using the approximation
(\ref{eq_ffkm:approximation_modelselection}) described in Section
\ref{sec_ffkm:model_selection}, the optimization problem of the FFKM
method with basis function expansions of the reduced functional data is
defined as
\begin{equation}
 \min_{\mathbf{A}_{H},U}\|\mathbf{F}_{pca}\mathbf{B}_{H}'\mathbf{A}_{H}
	-\mathbf{P}_{U}\mathbf{F}_{pca}\mathbf{B}_{H}'\mathbf{A}_{H}\|^{2}.
	\label{eq_ffkm:appendix_optimization1}
\end{equation}
We can see that $\mathbf{F}_{pca}\mathbf{B}_{H}'$ corresponds to
$\mathbf{G}_{H}$, which is the coefficient matrix of the basis function
expansion of the reduced functional data. Clearly, the rank of
$\mathbf{F}_{pca}\mathbf{B}_{H}'$ is $R$, i.e., the coefficient matrix
is rank deficient. Here, according to the recommendation by Vichi and
Kiers (2001), we consider eliminating the trivial dimensions of the
coefficient matrix. In the case of FFKM analysis, we can use
$\mathbf{F}_{pca}$ as the full-rank (neither singular nor near-singular)
matrix to be analyzed.

Therefore, instead of the optimization problem in
Eq. (\ref{eq_ffkm:appendix_optimization1}), the following optimization
problem is considered,
\begin{equation}
 \min_{\mathbf{A}_{H}^{*},U}\|\mathbf{F}_{pca}\mathbf{A}_{H}^{*}
	-\mathbf{P}_{U}\mathbf{F}_{pca}\mathbf{A}_{H}^{*}\|^{2},
	\label{eq_ffkm:appendix_optimization2}
\end{equation}
where $\mathbf{A}_{H}^{*}$ is an $R\times L$ orthogonal matrix that
spans an optimal subspace for representing the cluster structure. This
optimization problem can be solved by the same algorithm described in
Section \ref{sec_ffkm:algorithm}. That is, we just have to use
$\mathbf{F}_{pca}$ and $\mathbf{A}_{H}^{*}$ as $\mathbf{G}_{H}$ and
$\mathbf{A}_{H}$, respectively, in the algorithm for the FFKM method.

Using this two-step approach, we can obtain the cluster structure in the
low-dimensional subspace. However, this procedure does not provide the
weight functions $v_{l}$ that span the subspace of the functional
data. The weight functions are often useful to interpret the estimated
subspace and cluster structure. Thus, we consider estimating the weight
functions from the estimates $\mathbf{A}_{H}^{*}$.

To obtain the coefficient matrix $\mathbf{A}_{H}$ of the weight
functions $v_{l}$, the following optimization problem is considered,
\begin{equation}
 \min_{\mathbf{A}_{H}}\|\mathbf{F}_{pca}\mathbf{B}_{H}'\mathbf{A}_{H}
	-\mathbf{F}_{pca}\mathbf{A}_{H}^{*}\|^{2}.
\end{equation}
Note that $\mathbf{A}_{H}$ is an orthogonal matrix. This is the
well-known orthogonal Procrustes rotation problem (ten Berge, 1993), and
it can be solved easily. The singular value decomposition
$\mathbf{B}_{H}\mathbf{F}_{pca}'\mathbf{F}_{pca}\mathbf{A}_{H}^{*}=\mathbf{PDQ}'$
yields $\mathbf{A}_{H}=\mathbf{PQ}'$ as the optimizing solution, where
$\mathbf{P}'\mathbf{P}=\mathbf{Q}'\mathbf{Q}=\mathbf{I}_{L}$ and
$\mathbf{D}$ is a diagonal matrix whose diagonal element is a singular
value. Then, using
$\mathbf{A}_{H}=(\boldsymbol{a}_{H1},\dots,\boldsymbol{a}_{HL})$, the
estimated weight function is calculated as
$v_{l}=\boldsymbol{a}_{Hl}'\mathbf{H}^{-1/2}\boldsymbol{\phi}$. Furthermore,
we can obtain the component score matrix $\mathbf{F}$ as
$\mathbf{F}=\mathbf{G}_{H}\mathbf{A}_{H}$.

\end{document}